\documentclass[aps,prb,amsmath,amssymb,superscriptaddress,twocolumn]{revtex4-1}
\usepackage{graphicx}
\usepackage{dcolumn}
\usepackage{bm}
\usepackage{rotating}
\usepackage{multirow}
\usepackage[table]{xcolor}
\usepackage[english]{babel}
\usepackage{amssymb} 
\usepackage{amsmath} 
\usepackage[utf8]{inputenc} 
\usepackage{siunitx}
\newcommand{\be}{\begin{equation}}
\newcommand{\ee}{\end{equation}}
\newcommand{\bea}{\begin{eqnarray}}
\newcommand{\eea}{\end{eqnarray}}
\newcommand{\nn} {\nonumber}

\newcommand{\Tr}{ {\rm Tr} \, }

\def\a{\alpha}

\def\G{\Gamma}
\def\d{\delta}

\def\l{\lambda}

\def\S{\Sigma}

\def\w{\omega}

\def\xc{{\rm xc}}
\def\x{{\rm x}}

\def\Tr{{\rm Tr}\,}

\def\br{\mbox{\boldmath $r$}}

\begin{document}
\widetext 

\title{Beyond the random phase approximation with a local exchange vertex}
\author{Maria Hellgren}
\affiliation{Sorbonne Universit\'e, Mus\'eum National d'Histoire Naturelle, UMR CNRS 7590, IRD, Institut de Min\'eralogie, de Physique des Mat\'eriaux et de Cosmochimie, IMPMC, 4 place Jussieu, 75005 Paris, France}
\author{Nicola Colonna}
\affiliation{Theory and Simulation of Materials (THEOS), \'Ecole Polytechnique Federale de Lausanne, 1015 Lausanne, Switzerland}
\author{Stefano de Gironcoli}
\affiliation{International School for Advanced Studies (SISSA), Via Bonomea 265, I-34136 Trieste, Italy}
 \affiliation{CRS Democritos, CNR-IOM Democritos, Via Bonomea 265, I-34136 Trieste, Italy}
\date{\today}
\begin{abstract}
With the aim of constructing an electronic structure approach that systematically goes beyond the GW and random phase approximation (RPA) we introduce a vertex correction based on the exact-exchange (EXX) potential of time-dependent density functional theory. The EXX vertex function is constrained to be local but is expected 
to capture similar physics as the Hartree-Fock vertex. With the EXX vertex we then 
consistently unify different beyond-RPA approaches such as the various 
re-summations of RPA with exchange (RPAx) and the second order screened exchange (SOSEX) approximation. The theoretical analysis is supported 
by numerical studies on the hydrogen dimer and the electron gas, and we discuss the importance of including the vertex correction in both the screened interaction and the self energy.
Finally, we give details on our implementation within the plane-wave pseudo potential framework and 
demonstrate the excellent performance of the different RPAx methods in describing the energetics 
of hydrogen and van der Waals bonds. 
\end{abstract}
\maketitle

\section{Introduction} 
Kohn-Sham (KS) density functional theory (DFT)\cite{hohenbergkohn,kohnsham1965} has 
been extremely successful in predicting structural and vibrational properties of 
a wide range of systems. However, for sparse systems or systems in reduced dimensions,
the standard local or semi-local approximations such as LDA and GGAs are unreliable 
and sometimes fail qualitatively. Examples are the various silica polymorphs 
which require a proper description of the van der Waals forces,\cite{henri2015,zeolites} and the layered 
materials which often exhibit effects of strong electron correlation.
Moreover, in reduced dimensions, excitonic effects become increasingly important 
and may be strong enough to influence ground-state properties.\cite{nanoexc} 

Although highly accurate methods such as coupled cluster or quantum Monte Carlo (QMC)
could overcome these limitations high computational cost limits them 
to the study of very selected systems. A cheaper solution to at least some of the problems 
is to include a fraction of Hartree-Fock (HF) exchange via the so-called hybrid 
functionals.\cite{hse06} These functionals are well-known to improve the electronic structure of 
systems containing, e.g., transition metal atoms. On the other hand, the hybrid 
functionals miss the van der Waals forces and rely on a parametrisation which 
often turns out to be system dependent. 

Another route is to use many-body perturbation theory (MBPT) based on Green's functions
which is a complete theoretical framework for studying both ground and excited states properties. 
A popular approximation to the many-body self-energy $\S$ is the GW approximation\cite{hedin} (GWA), 
which simultaneously captures the exchange interaction and the van der Waals forces. 
The GWA is widely used for band-structure calculations (i.e. quasi-particle energies) and 
total energy calculations are becoming feasible, at least within the random phase approximation 
(RPA), which can be seen as the variationally best 
GW total energy within a restricted space of KS Green's functions.\cite{dahlenvar,dahlenbarth06,hvb07}
Approximations beyond the GW self-energy can be constructed using diagrammatic techniques by, 
for example, including the second order screened exchange (SOSEX) diagram.
\cite{freeman,doi:10.1063/1.3250347,PhysRevB.88.035120,PhysRevB.92.081104,PhysRevB.47.15404} 
Since, in general, it is not obvious which set of diagrams to include different approximation schemes 
have been developed. In the so-called $\Phi/\Psi$-derivable schemes\cite{almbladh} diagrams are chosen 
such that only conserving approximations are generated, i.e., approximations that conserve energy, 
momentum and particle number.\cite{baym,baymkadanoff} Another approach is Hedin's 
iterative procedure, in which approximations are generated from a vertex function $\G=1+\d\S/\d V$ 
where $\S=iGW\G$ and $V$ is the single particle potential.\cite{hedin} By, e.g., starting from 
GW (i.e. $\G=1$) the self-energy can be updated iteratively. It is also possible to 
start from a simpler approximation to the self-energy, such as the HF approximation, 
generating a non-local vertex function similar to the one used in the Bethe-Salpeter equation. 
 
In this work we study an approximate local HF vertex derived from the exact-exchange (EXX)
potential of time-dependent density functional theory (TDDFT). The EXX potential is 
obtained by minimizing the HF energy in a constrained space of local potentials.\cite{oep}
Local vertex corrections derived from LDA and GGA's have been studied in the context of quasi-particle energies, but their effect on the GW gaps was quite small.\cite{PhysRevB.49.8024,PhysRevLett.94.186402,doi:10.1063/1.3249965} The HF or the 
local EXX vertex should more accurately capture the electron-hole interaction, expected 
to be important for correcting the deficiencies found within the self-consistent GWA.\cite{PhysRevB.95.035139}
The main focus of this paper will be on the total energy and we will show how
one can consistently generate different approximations beyond the RPA total energy 
using the EXX vertex. In this way, we establish theoretical connections between the
different variants of RPA with exchange (RPAx), usually defined in terms of the density response function,
\cite{toulouse_prl,hvb09,hesselmann_random_2010,hvb10,hg11,RPAx-F,PhysRevB.90.125150,dixit_improving_2016,mussard_dielectric_2016,dixit17} 
as well as between the RPAx and the SOSEX total energy approximation.
The analysis also provides a many-body perspective on approximations usually defined within DFT. 
 
The paper is organized as follows. In Sec. II we derive the basic equations for the 
adiabatic connection total energy formula within MBPT and DFT. We then define the vertex function,
which establishes a connection between the two frameworks. We also review the comparison between 
GW and RPA. In Sec. III we define the local HF or EXX vertex and use it to derive 
previously defined approximations based on RPAx and a slightly new variant of SOSEX.
In Sec. IV we present numerical results on H$_2$ and the electron gas to
support the theoretical discussion in Sec. III. We also present results on the A24 test-set~\cite{rezac_describing_2013}
to demonstrate the performance of the different RPAx methods in describing hydrogen and van der Waals bonds. In the end, we present numerical details from our implementation within the plane-wave and pseudo potential framework. Finally, in Sec. V, we present our conclusions.

\section{Correlation energy}
\label{sec:corr_ene}
We start by deriving the exact adiabatic connection formula
for the correlation energy in terms of the single-particle Green's function $G$. 
Similar derivations can be found in Refs.~\onlinecite{fetter,langreth}.

The Hamiltonian of the interacting electronic system is given by
\be
\hat H = \hat T+\hat V_{\rm ext}+\hat W
\ee
where $\hat T$ is the kinetic energy operator, $\hat V_{\rm ext}$ is the external 
nuclear potential and $\hat W$ is Coulomb interaction between the electrons.
The adiabatic connection path is defined in terms of a parameter $\l\in[0,1]$ 
that linearly scales the Coulomb interaction $\hat W$ from zero to full interaction 
strength. A single-particle potential $V^{\l}$ is added to the Hamiltonian such 
that the density is kept fixed to its interacting value ($\l=1$) at every value 
of $\l$. Introducing the Hartree (H) and exchange-correlation (xc) potential $\hat V_{\rm Hxc}$, 
as defined within KS DFT, we thus write the scaled Hamiltonian on the adiabatic 
connection path as follows
\be
\hat H^\l = \hat T+\hat V_{\rm ext}+\hat V_{\rm Hxc}-\hat V^{\l}+\l\hat W
\label{scham}
\ee
where
\be
\hat V^{\l=1}=\hat V_{\rm Hxc},\,\,\,\,\hat V^{\l=0}=0.
\ee
With the Hamiltonian defined in Eq.~(\ref{scham}) we derive the corresponding 
single-particle Green's function and write it in terms of a Dyson equation 
\be
G_{\lambda}=G_s+G_s[\Sigma_{\lambda}[G_{\lambda}]-V^{\lambda}]G_{\lambda}
\ee
where $\S_{\lambda}$ is the irreducible self-energy and $G_s$ is the Green's function 
of the $\l=0$ system (i.e. the KS Green's function). Using standard tricks\cite{fetter} 
we can write the total energy at full interaction strength as an integral over 
the coupling constant $\l$. Defining the single-particle energy 
\be
E_s=T_{s}+\int nv_{\rm ext}+E_{\rm H}
\ee
the total energy is written as
\be
E=E_s-\frac{i}{2}\int \frac{d\w}{2\pi}\int \frac{d\lambda}{\lambda}{\rm Tr}\{\Sigma_{\lambda}[G_{\lambda}]G_{\lambda}\}
\label{eself}
\ee
where the trace is defined as $\Tr\{ AB \}=\int d{\bf r} d{\bf r}' A({\bf r},{\bf r}')B({\bf r}',{\bf r})$ and the convergence factor ($e^{i\w\eta}$) has been suppressed. 
This is an exact formula from which one can generate approximate total energies from any 
approximate $\S$.

An alternative way to write the correlation energy is in terms of the density correlation function 
(or the reducible polarization propagator) $\chi$
\be
E=E_s +\frac{i}{2}\int \frac{d\w}{2\pi}\int\! d\lambda\,{\rm Tr}\{v[\chi_{\lambda}-\delta n]\}
\label{echi}
\ee
where $\delta n=\d({\bf r},{\bf r}')n({\bf r})$. In this form, approximations are generated diagrammatically 
via the Bethe-Salpeter equation, or via the linear response TDDFT Dyson equation. 

In order to connect the two expressions, Eq.~(\ref{eself}) and Eq.~(\ref{echi}), we define the irreducible polarization propagator $P$ in terms of the vertex function $\G$ 
\be
P = -iGG\G,\,\,\,\,\G=-\frac{\d G^{-1}}{\d V}=1+\frac{\d \S}{\d V}
\label{defvert}
\ee
where $V=v_{\rm ext} + v_{\rm H}$. With these quantities it is possible to rewrite the 
scaled density correlation function using the relation 
\be
\chi_\l v=\frac{1}{\l}P_\l W_\l
\ee
in which $W=v+vPW$ is the screened Coulomb interaction. Since $P$ is 
given by Eq.~(\ref{defvert}) and $\S=iG W\G$ we can rewrite Eq.~(\ref{echi}) in the form of Eq.~(\ref{eself}).

Within TDDFT the density correlation function is determined from the time-dependent 
KS system which is defined to reproduce the exact interacting density with a time-dependent 
local single-particle potential. The TDDFT Dyson equation reads
\be
\chi_\l=\chi_s+\chi_s[\l v+f^\l_\xc]\chi_\l
\label{dystddft}
\ee 
where $\chi_s=-iG_sG_s$ is the KS polarization propagator and $f^\l_\xc$ is the functional 
derivative of the time-dependent xc potential with respect to the density. 
The irreducible polarization propagator $P$ is then determined
from $P=\chi_s+\chi_sf_{\xc}P$. But, this expression can easily be rearranged to 
$P=-iG_sG_s\G_{\xc}$, that is, in terms of a local vertex function defined as
\be
\G_\xc=1+\frac{\d v_{\xc}}{\d V}.
\ee
In this way we can rewrite Eqs.~(\ref{echi}) as 
\be
E=E_s-\frac{i}{2}\int \frac{d\w}{2\pi}\int \frac{d\lambda}{\lambda}{\rm Tr}\{\Sigma^{GW\G_{\xc}}_{\lambda}G_s\}
\label{eselflocal}
\ee
where $\Sigma^{GW\G_{\xc}}=iG_sW\G_\xc$, which can be compared to Eq.~(\ref{eself}). 
Notice that if the exact xc kernel is used this is an exact expression for the correlation energy. 
This, however, does not mean that $\Sigma^{GW\G_{\xc}}$ is the exact self-energy which
should be defined in terms of the fully non-local vertex function. The extent to which a local vertex
can approximate the exact vertex was studied in 
Ref.~\onlinecite{doi:10.1063/1.3249965}.

\subsection{RPA and GW}
The GWA and RPA have already been thoroughly compared\cite{gwrpa1,gwrpa2} but for the 
sake of completeness we review this comparison here. Within the GWA we can 
calculate the total energy via Eq.~(\ref{eself}) by simply setting 
\be
\S^{\rm GW}_{\lambda}=iG_\l W_\l,\,\,\,\, W_\l=\l v+\l vP_0W_\l.
\ee
The screened interaction is calculated in the time-dependent Hartree approximation (or RPA)
for which $P_0=-iGG$. 

The RPA for the total energy is given by 
\be
E^{\rm RPA}=E_{s} +\frac{i}{2}\int \frac{d\w}{2\pi}\int d\lambda{\rm Tr}\{v[\chi^{\rm RPA}_{\lambda}-\delta n]\}
\label{erpachi}
\ee 
where
\be
\chi_\l^{\rm RPA}=\chi_s+\l\chi_sv\chi_\l^{\rm RPA}.
\ee 
Following the steps in the previous section [Eqs.~(\ref{defvert})-(\ref{eselflocal})] we see that 
this response function corresponds to $\G_\xc=1$. We can thus write 
\be
E^{\rm RPA}=E_s-\frac{i}{2}\int \frac{d\w}{2\pi}\int \frac{d\lambda}{\lambda}{\rm Tr}\{\Sigma^{\rm GW}_{\lambda}[G_s]G_s\}
\label{eq:dyson_rpa}
\ee
and we see that this exactly corresponds to Eq.~(\ref{eself}) in the GWA but with $G_\l$ replaced by
$G_s$. We notice that the diagrammatic structure within the GWA is preserved in the RPA. Furthermore, 
since only the explicit $\l$-dependence is retained, the $\l$-integral is easily carried out leading to the 
following expression
\be
E^{\rm RPA}=E_s-\frac{i}{2}\int \!\frac{d\w}{2\pi}{\rm Tr}\{\ln [1+ivG_sG_s]\},
\label{rpa}
\ee
where we have suppressed the frequency integral.
Since the GW self-energy is $\Phi$-derivable, meaning that the self-energy can be generated from a 
functional $\Phi[G]$, i.e., $\S=\d\Phi[G]/\d G$, also the $\l$ integral in Eq.~(\ref{eself}) can be 
performed analytically. This leads to a total energy of the form
\bea
E^{\rm GW}&=&-\frac{i}{2}\int \!\frac{d\w}{2\pi}{\rm Tr}\{\ln [1+ivGG]\}+E_{\rm H}\nn\\
&&\,\,\,\,\,\,\,\,\,+\,i\int\! \frac{d\w}{2\pi}\Tr [GG_s^{-1}-1+\ln (-G^{-1})].
\eea
This is a functional of $G$ (also known as the Klein functional) which is stationary when $G$ 
obeys the Dyson equation. If we replace $G$ with $G_s$ it is easy to 
show that this expression is equivalent to Eq.~(\ref{rpa}). We thus have a second way to 
compare the GWA and RPA.\cite{gwrpa1,gwrpa2} The RPA is the GW Klein energy constrained to the space of 
KS Green's functions. 

\section{Local Hartree-Fock vertex}
\label{sec:HF_vertex}
\begin{figure}[t]
\includegraphics[scale=0.3]{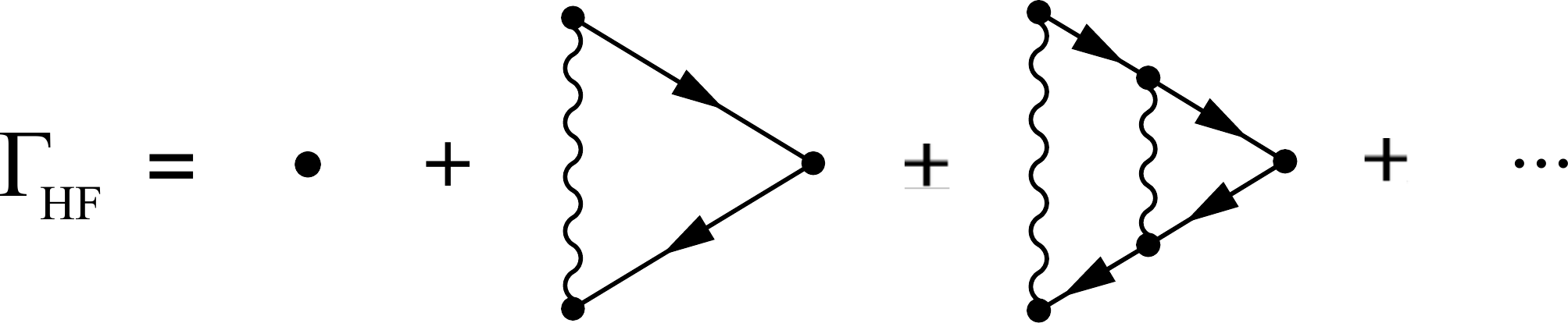}
\caption{Diagrammatic expansion of the vertex in the HF approximation.}
\label{gammahf}
\end{figure}

We will now improve the GWA and RPA by constructing an approximate vertex function 
(Eq.~(\ref{defvert})) from the HF self-energy ($\S_{\rm HF}=iGv$)
\be
\G_{\rm HF}(123)=\d (12)\d (13) + \frac{\d \S_{\rm HF}(12)}{\d V(3)}.
\ee
Here we have used the notation $1=\br_1,t_1$ to demonstrate that the vertex depends on 
three space and time variables. The diagrammatic expansion of the HF vertex is given 
in Fig. \ref{gammahf}. From the diagrammatic structure it is easy to see that the 
electron-hole interaction is accounted for. If we replace the bare Coulomb interaction with a statically 
screened Coulomb interaction this vertex generates what is usually referred to as 
the Bethe-Salpeter equation for the polarization propagator.

\begin{figure}[b]
\includegraphics[scale=0.28]{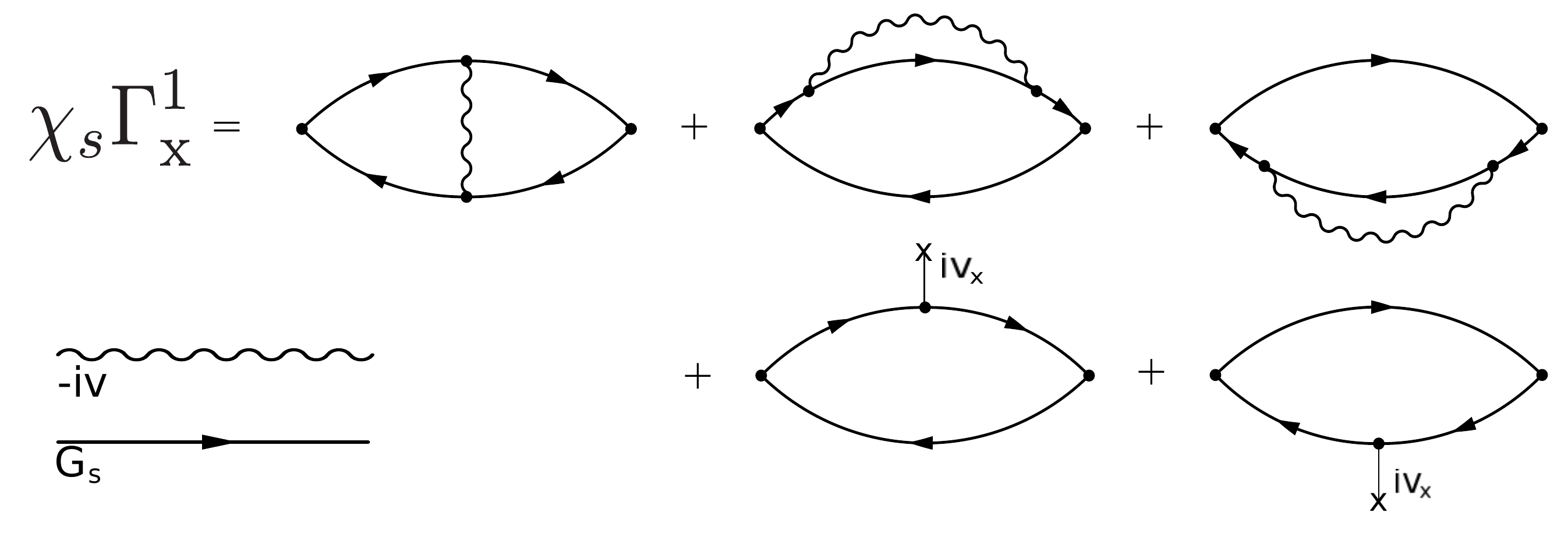}
\caption{Equation to obtain the first order local HF vertex $\G^1_{\x}$.}
\label{localvert}
\end{figure}

The corresponding local HF vertex can be defined in terms of the local time-dependent
exact-exchange (EXX) potential $v_\x$ 
\be
\G_{\rm x}= 1 + \frac{\d v_{\rm x}}{\d V}=\frac{1}{1-\G_\x^1}
\label{defexxvert}
\ee
where
\be
\G_\x^1=\frac{\d v_\x}{\d V_s},\,\,\,\,V_s=v_{\rm ext}+v_{\rm Hx}.
\label{vert1}
\ee
The time-dependent EXX potential is determined by the linearized Sham-Schl\"uter equation 
and has been studied in several previous work. It can, e.g., be shown to exactly 
reproduce the HF density to first order in the Coulomb interaction. 
The equation for $\G_\x^1$ is illustrated diagrammatically 
in Fig. \ref{localvert} and we see that the local EXX vertex has a similar structure to the full HF vertex. Using the chain rule, Eq.~(\ref{vert1}) also defines the standard EXX kernel 
\be
f_\x=\frac{\d v_\x}{\d n}=\G_\x^1 \chi_s^{-1}.
\ee
With the EXX kernel we can generate the TDDFT density correlation function which is usually 
called the RPAx response function
\be
\chi^{\rm RPAx}=\chi_s+\chi_s[ v+f_\x]\chi^{\rm RPAx}.
\label{respx}
\ee
This response function has been applied to atoms and molecules as well as to silicon for the calculation of excitons.\cite{hvb08,hvb09,hvb10,kg02} Although some excitation energies are sensitive to the local approximation\cite{hvb09} integrated quantities such as polarizabilities and van der Waals coefficients give 
results similar to the nonlocal time-dependent HF (TDHF) approach.\cite{hvb10} 
In the following we will 
use the EXX vertex to generate a set of approximations to the total energy.

\subsection{GW$\G_{\rm x}$ and GW$\G_{\rm HF}^1$}
If we use the RPAx response function defined in Eq.~(\ref{respx}) and insert it into the total 
energy of Eq.~(\ref{echi}) we find what we have previously called the RPAx for the total energy
\be
E^{\rm RPAx}=E_{s} +\frac{i}{2}\int \frac{d\w}{2\pi}\int d\lambda{\rm Tr}\{v[\chi^{\rm RPAx}_{\lambda}-\delta n]\}.
\label{erpaxchi}
\ee 
With the definition of the EXX vertex in Eq.~(\ref{defexxvert}) we can follow the same steps that lead 
to Eq.~(\ref{eselflocal}) and rewrite Eq.~(\ref{erpaxchi}) as 
\be
E^{{\rm RPAx}}=E_{s}-\frac{i}{2}\int \frac{d\w}{2\pi}\int \frac{d\lambda}{\lambda}{\rm Tr}\{\Sigma^{\rm GW\G_{\rm x}}_{\lambda}G_s\}
\ee
with $\Sigma^{GW\G_{\rm x}}=iG_sW\G_{\rm x}$. Since the vertex is local this self-energy is approximate 
but is expected to mimic the self-energy with the full HF vertex. If the latter is used, the approximation 
to the correlation energy corresponds to the RPAx defined in Ref.~\onlinecite{toulouse_prl}, 
i.e., the correlation energy with the TDHF response function. 
Comparing the results of Refs.~\onlinecite{toulouse_prl,toulousejcp} to those of 
Ref.~\onlinecite{PhysRevB.93.195108} we see similar trends. Only in the case of the Be dimer there is a qualitative difference and it is found that the use of a nonlocal potential for generating $G_s$ strongly improves the results. We notice that neither the HF nor the EXX vertex produces $\Phi$-derivable self-energies due to the
lack of symmetry. 

If we expand the EXX vertex to first order only ($\G_{\rm x}^1$) all terms in the expansion of the 
correlation energy have a diagrammatic representation. We can therefore write the total energy of 
the $\G_{\rm x}^1$-approximation [RPAx(1)] in terms of the full HF vertex to first order 
($\G^{1}_{\rm HF}$) plus a self-energy correction 
\be
E^{{\rm RPAx}(1)}=E_{s}-\frac{i}{2}\int \frac{d\w}{2\pi}\int \frac{d\lambda}{\lambda}{\rm Tr}\{\Sigma^{\rm GW\G^{1}_{\rm HF}}_{\lambda}[G_s]G_s\}+E_{\S_s}.
\label{rpax1}
\ee
The diagrammatic representation of the first correlation term is given in Fig.~\ref{sosex}(a). 
The second, self-energy correction, is of the form
\be
E_{\S_s}=-\frac{i}{2}\int \frac{d\w}{2\pi}\int d\lambda\,{\rm Tr}\{G_sW^{\l}G^1+G^1 W^{\l}G_s\}
\label{sigself}
\ee
with
\be
G^1=G_s[\S_{\rm HF}-v_\x]G_s.
\ee
One of the terms in Eq.~(\ref{sigself}) is illustrated in Fig.~\ref{sosex}(c). 

Alternatively we can write the correlation energy of Eq.~(\ref{rpax1}) as in Eq.~(\ref{echi}), in terms 
of an approximate density correlation function obtained by expanding the irreducible 
polarizability to first order
\bea
\chi^{\rm RPAx(1)}_{\lambda}&=&P^1_\lambda+\l P^1_\lambda v\chi^{\rm RPAx(1)}_{\lambda}\nn\\
&&\Rightarrow \chi^{\rm RPAx(1)}_{\lambda}=[1-\l P^1_\lambda v]^{-1} P^1_\lambda
\label{eq:dyson_rpax1}
\eea
where $P^1_\lambda=\chi_s+\lambda\chi_s\G^1_{\rm x}=\chi_s+\lambda\chi_sf_{\rm x}\chi_s$. This approximation was studied 
previously for the electron gas in Ref.~\onlinecite{PhysRevB.90.125150} and for the total energy of molecules in Ref.~\onlinecite{PhysRevB.93.195108}. By keeping the vertex to first order only cured certain pathologies due to unscreened Coulomb interaction.

\subsection{SOSEX}
\begin{figure}[t]
\includegraphics[scale=0.6]{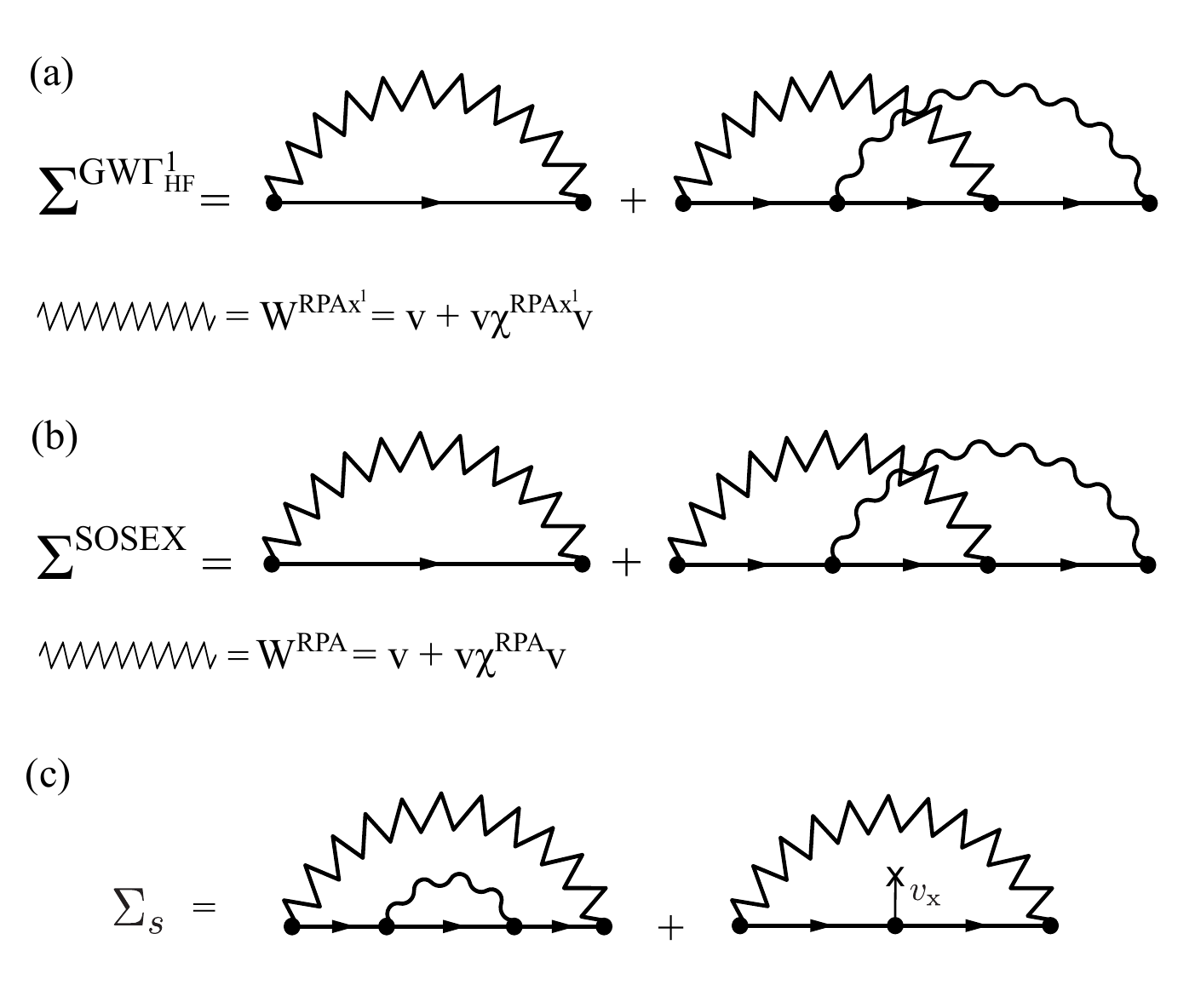}
\caption{Diagrammatic representation of the $GW\G_{\rm HF}^1$ (a) and SOSEX (b) self-energies. The self-energy correction in Eq.~(\ref{sigself}) is illustrated in (c).}
\label{sosex}
\end{figure}

In the RPAx(1) approximation defined above the vertex is treated to first order in both screened interaction 
and in the self-energy. If we set the vertex to one in the screened interaction (i.e. keeping it at the RPA level)
we generate the so-called SOSEX approximation. For a diagrammatic representation see Fig. \ref{sosex}(b). 
The total energy is given by
\bea
E^{{\rm AC-SOSEX}}&=&\nn\\
&&\!\!\!\!\!\!\!\!\!\!\!\!\!\!\!\!\!\!\!\!\!\!\!\!\!\!\!\!\!\!\!\!\!\!E_s-\frac{i}{2}\int \frac{d\w}{2\pi}\int \frac{d\lambda}{\lambda}{\rm Tr}\{\Sigma^{\rm SOSEX}_{\lambda}[G_s]G_s\}+\tilde E_{\S_s}
\eea
where $\tilde E_{\S}$ is a self-energy correction with an RPA screened interaction. This expression can
also be exactly rewritten in terms of an approximate density correlation function given by
\bea
\chi^{\rm SOSEX}_{\lambda}&=&P^1_\lambda+\l\chi_sv\chi^{\rm SOSEX}_{\lambda}\nn\\
&&\Rightarrow \chi^{\rm SOSEX}_{\lambda}=[1-\l\chi_sv]^{-1}P^1_\lambda
\label{eq:dyson_sosex}
\eea
The SOSEX has been studied previously, both for total energies, and quasi-particle excitations
of finite systems.\cite{freeman,doi:10.1063/1.3250347,PhysRevB.88.035120,PhysRevB.92.081104} 
Total energies are obtained either from the adiabatic connection expression (Eq.~(\ref{eself})) or from the perturbative 
Galitskii-Migdal expression. It has been shown that so-called single-excitations improve the results\cite{PhysRevB.88.035120} 
and our self-energy correction $\tilde E_{\S_s}$ can be interpreted as an approximate single-excitation 
correction. In our formalism this correction naturally arises from the definition of the local EXX vertex. 
The so generated set of approximations RPAx, RPAx(1) and AC-SOSEX can thus be seen as a consistent way to do many-body approximations based on a local DFT framework. 
Self-energy corrections to the same order as the vertex naturally follows and 
compensate to some extent for a lack of self-consistency within the MBPT scheme. 
Once the EXX vertex (or the EXX kernel) is calculated all approximations follow straightforwardly (see also Appendix A). 
In the next section we will compare these different approximations on H$_2$ and the electron gas and assess their performance in describing hydrogen and van der Waals bonds.

\section{Numerical results}
In this section we present numerical results on H$_2$ and the 
electron gas, in order to compare the different approximations defined above, and to 
compare to previous SOSEX results in the literature. We also present calculations on 
the A24 test-set~\cite{rezac_describing_2013} containing molecules with hydrogen and van der Waals bonds. 
All calculations (except for those on the electron gas) have been carried out with a recently released RPA/RPAx-implementation 
within the {\sc Quantum Espresso} package.~\cite{giannozzi_advanced_2017} 
See Sec.~IVC, Appendix A and Refs.~\onlinecite{PhysRevB.90.125150,nguyen_efficient_2009,nguyen_ab_2014}
for details regarding the implementation.

\subsection{H$_2$ and the electron gas}
\begin{figure}[t]
\includegraphics[width=\columnwidth]{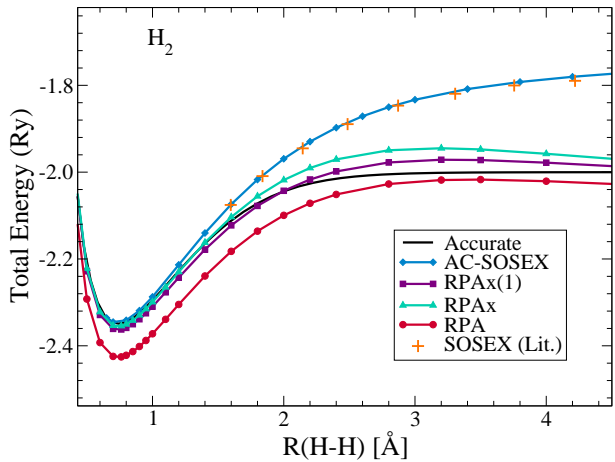}
\caption{Dissociation curves for the hydrogen molecule within different beyond-RPA methods. Approximations derived in this work are compared to 'exact' results presented in Ref.~\onlinecite{kolos} and to SOSEX results obtained by Ren {\it et al.}\ in Ref.~\onlinecite{Ren2012}}.
\label{h21}
\end{figure}

In Fig.~\ref{h21} we present H$_2$ dissociation curves within RPA, RPAx, RPAx(1) 
and AC-SOSEX, and compare the results to accurate results from Ref.~\onlinecite{kolos}. We also 
compare our AC-SOSEX to previous SOSEX results in the literature.\cite{Ren2012}
RPA, RPAx and RPAx(1) all consistently include the vertex correction in both 
the screened interaction and the self energy ($\G=1$, $\G_{\rm x}$ and 
$\G^1_{\rm HF}$ respectively). At the same time they are all able to 
capture the dissociation region, or static correlation, reasonably well.
Our AC-SOSEX results are essentially identical to previous results. SOSEX includes the vertex correction ($\G^1_{\rm HF}$) in the self-energy
only, while it treats the screened interaction at the RPA level. This inconsistency
could explain why it fails to describe dissociation into open-shell atoms.

\begin{figure}[t]
\includegraphics[width=\columnwidth]{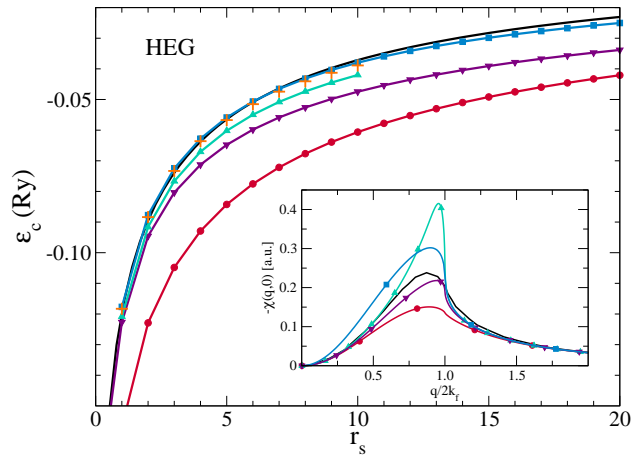}
\caption{The energy per particle of the homogeneous electron gas within different beyond-RPA methods compared to QMC results of Ref.~\onlinecite{egqmc} and to SOSEX results of Ref.~\onlinecite{freeman}.
Symbols and line coding as in fig.~\ref{h21}. The inset shows the corresponding static response functions.}
\label{h22}
\end{figure}

In Fig.~\ref{h22} we present the correlation energy per particle in the 
homogeneous electron gas (HEG). We compare our AC-SOSEX energies to previous
SOSEX results by Freeman\cite{freeman} and to accurate QMC results.\cite{ceperley_ground_1980,egqmc}
Again, we find that our AC-SOSEX agrees very well with previous SOSEX results.
Both coincides with QMC results at a given density (AC-SOSEX
at $r_s=6.3$ and SOSEX at $r_s=4.8$). Overall, SOSEX produces energies in very good agreement with QMC even up to low densities. RPAx(1) represents a systematic
improvement upon RPA, performing very well in the metallic range ($r_s<5$) although
slightly worse than RPAx. For a more detailed discussion of RPAx and RPAx(1), see 
Ref.~\onlinecite{PhysRevB.90.125150}.
But, we mention here that expanding the EXX vertex to first order only, 
i.e. going from RPAx to RPAx(1), removes an instability of the response function 
at low electronic densities~\cite{PhysRevB.90.125150} $(r_s>11)$. A similar 
behavior has been also reported by the author of Ref.~\onlinecite{maggio_correlation_2016}
who observed imaginary frequencies eigenmodes for the polarization propagator
computed solving the Bethe Salpeter Equation (BSE). Also in this case disregarding 
selected diagrams in the BSE kernel remove the instability in the low density regime. 

In the inset of  Fig.~\ref{h22} we have plotted the static response functions at $r_s=5$,
obtained from Eq.~(\ref{respx}), Eq.~(\ref{eq:dyson_rpax1}) and Eq.~(\ref{eq:dyson_sosex}).
Again, RPAx(1) systematically improves upon RPA and has no pathological behavior as compared
to RPAx. SOSEX strongly overestimates the static response up to its maximum 
suggesting that the good energies are subject to effects of error cancellation.

\subsection{Dispersion interactions: The A24 test-set}
\begin{figure*}[t]
\includegraphics[width=1.5\columnwidth]{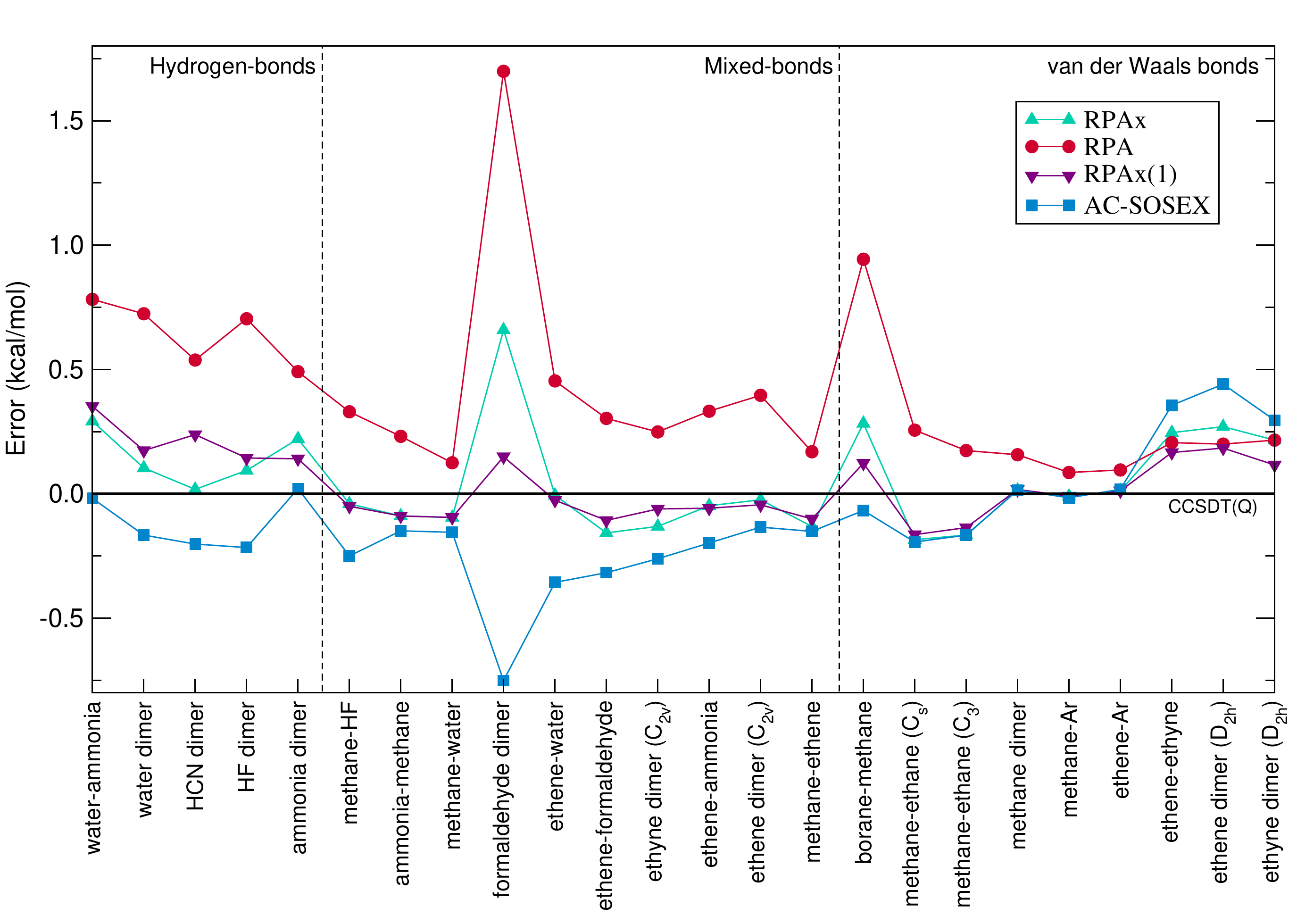}
\caption{Error in binding energies (kcal/mol) for the molecules in the A24 test-set. Results are compared to highly accurate CCSDT(Q) results of Ref.~\onlinecite{refA24}. Data can be found in Table 1.}
\label{a24}
\end{figure*}

In order to test the performance of the beyond-RPA methods in describing dispersion 
forces we calculated the binding energies of molecular dimers in the A24 test set.~\cite{rezac_describing_2013} 
The results are presented in Fig.~\ref{a24} and in Table 1. The RPA tends to underestimate
the binding energies with a Mean Absolute Error (MAE) of 0.44 kcal/mol. SOSEX has been shown to improve the description of dispersion interactions as compared to RPA and we find similar results in this work, with a MAE of 0.19 kcal/mol. Overall, the SOSEX energies are overestimated, in particular 
for the dimers with mixed bonds where the errors only slightly reduce with respect to RPA. 
RPAx and RPAx(1) give similar average performances 
(MAE of 0.14 kcal/mol and 0.13 kcal/mol, respectively) but RPAx(1) gives a more reliable improvement without exceptions. 

Comparing to other beyond-RPA results in the literature we find that our approximations perform better than those of TDHF based approximations, unless a range-separation parameter 
is introduced.\cite{toulousejcp,toulousejcp1,dixit17} As expected, our RPAx results agree with the RPAx results of Ref.~\onlinecite{prlerhard}.
\subsection{Technical aspects} 
We performed the molecular ACFDT calculations in a plane-wave and pseudo-potential formalism as implemented in a separate module of the {\sc Quantum ESPRESSO} distribution.~\cite{giannozzi_quantum_2009, giannozzi_advanced_2017}
We devote this section to review the basic aspects of our implementation of the exchange and correlation energy at all the levels of the theory discussed in Sec.~\ref{sec:corr_ene} and Sec.~\ref{sec:HF_vertex}
(for a more detailed description we refer to Refs.~\onlinecite{nguyen_efficient_2009,nguyen_ab_2014,PhysRevB.90.125150}).
We complement the discussion with convergence tests on the relevant parameters of our implementation for a subset composed of three complexes, namely the water-ammonia, the methane-HF and the borane-methane, each one representative for a different type of bond in the A24 set. 

\begin{figure*}
 \includegraphics[height=0.2\textwidth]{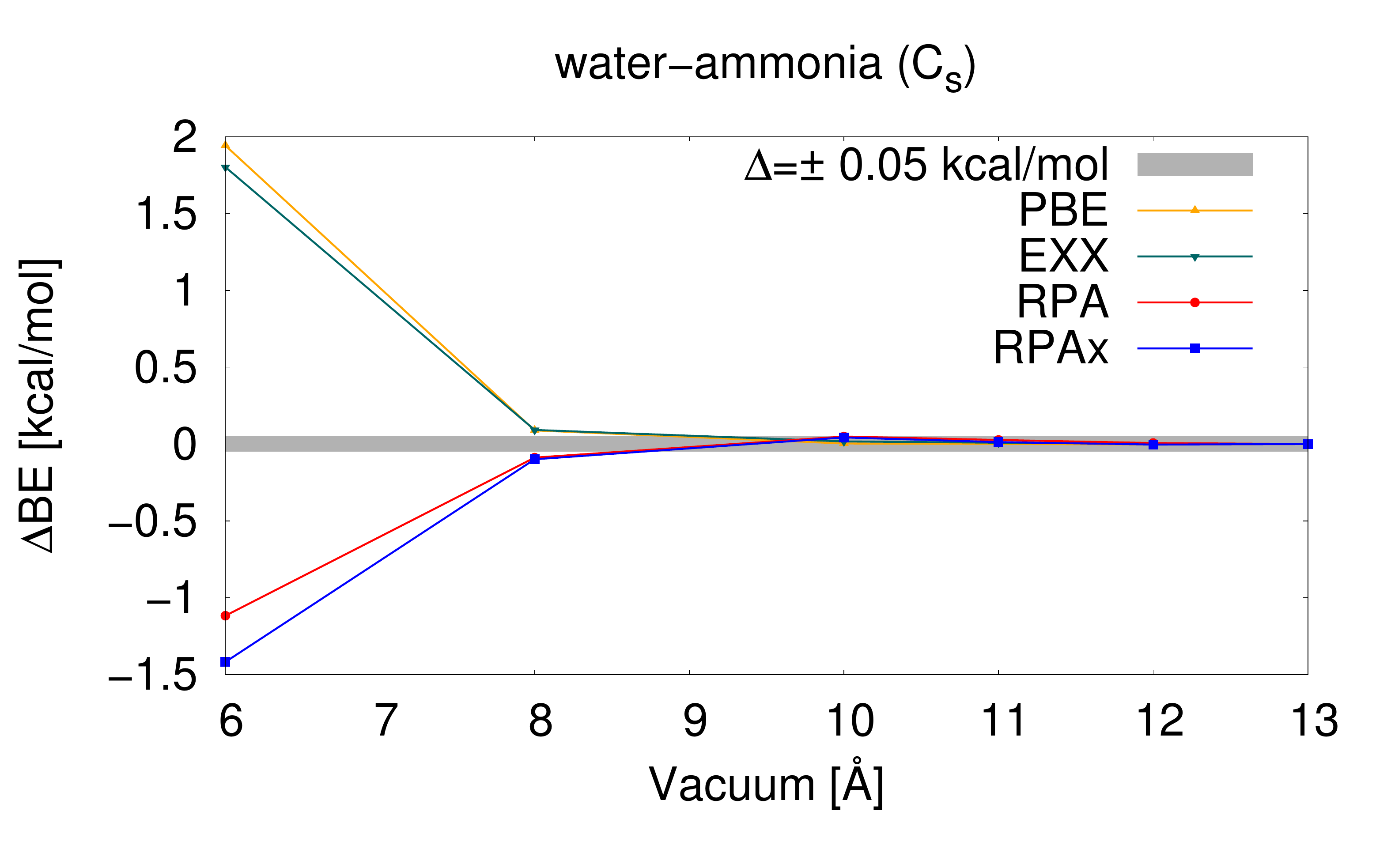}
 \includegraphics[height=0.2\textwidth]{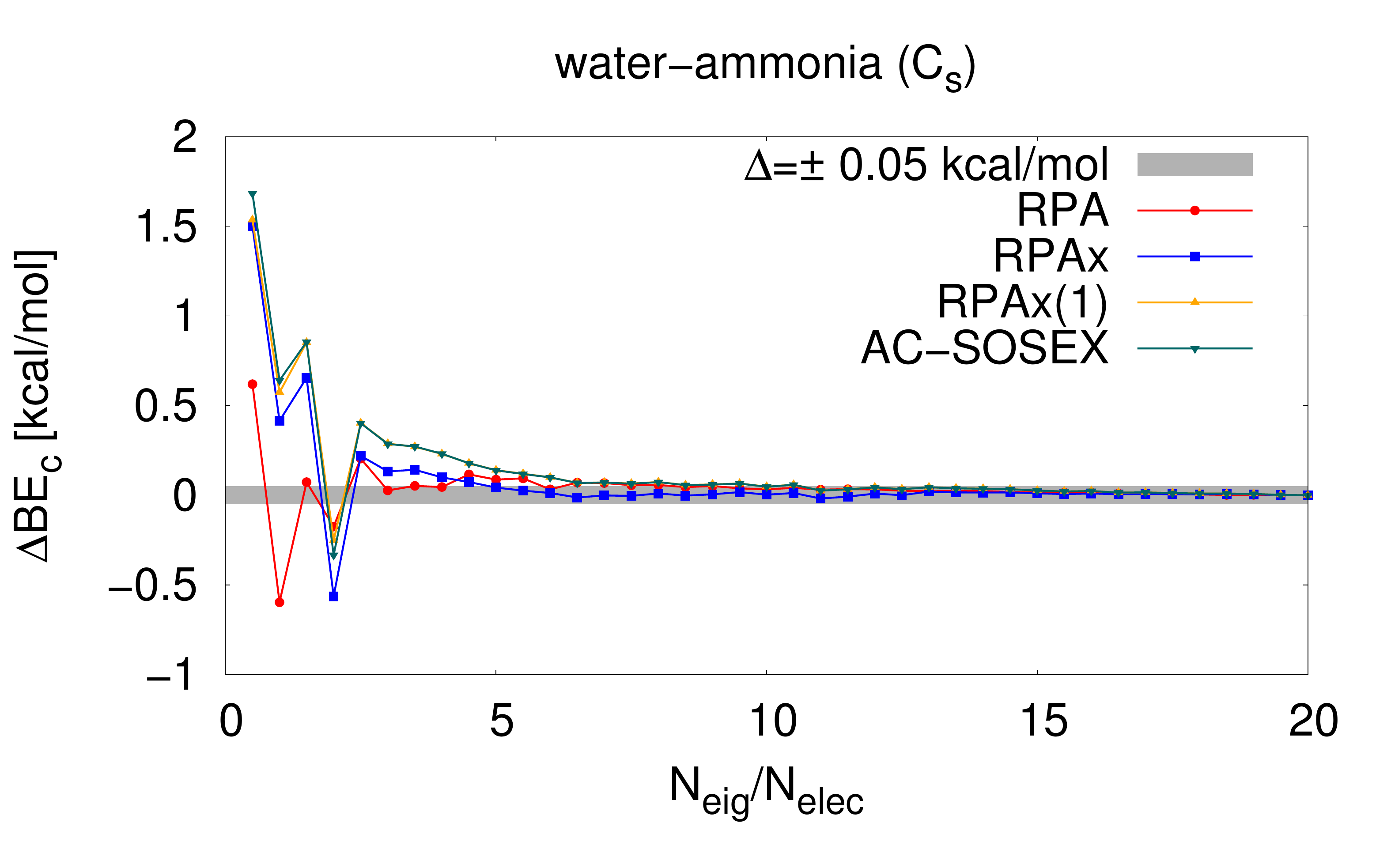}
 \includegraphics[height=0.2\textwidth]{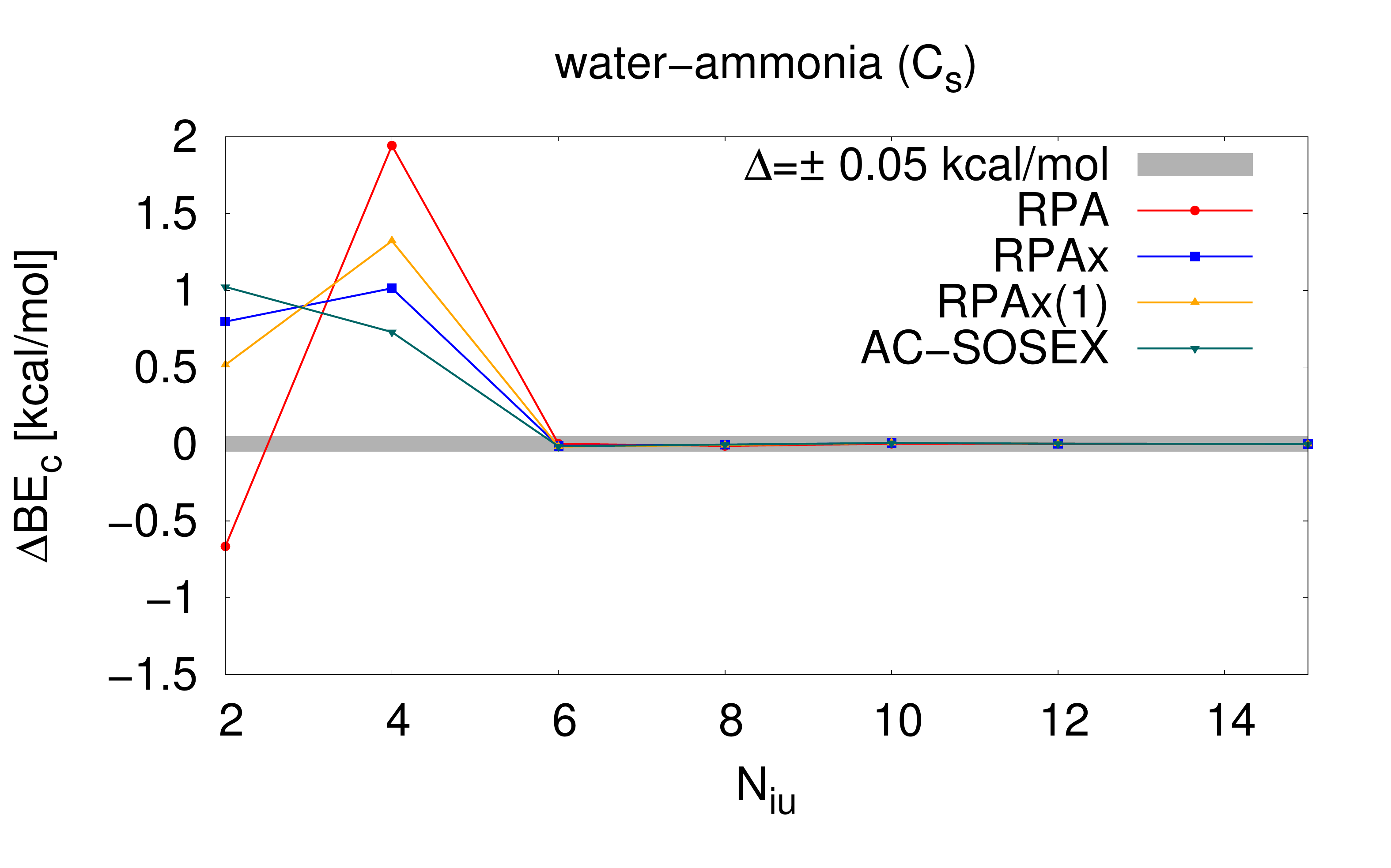}
 \includegraphics[height=0.2\textwidth]{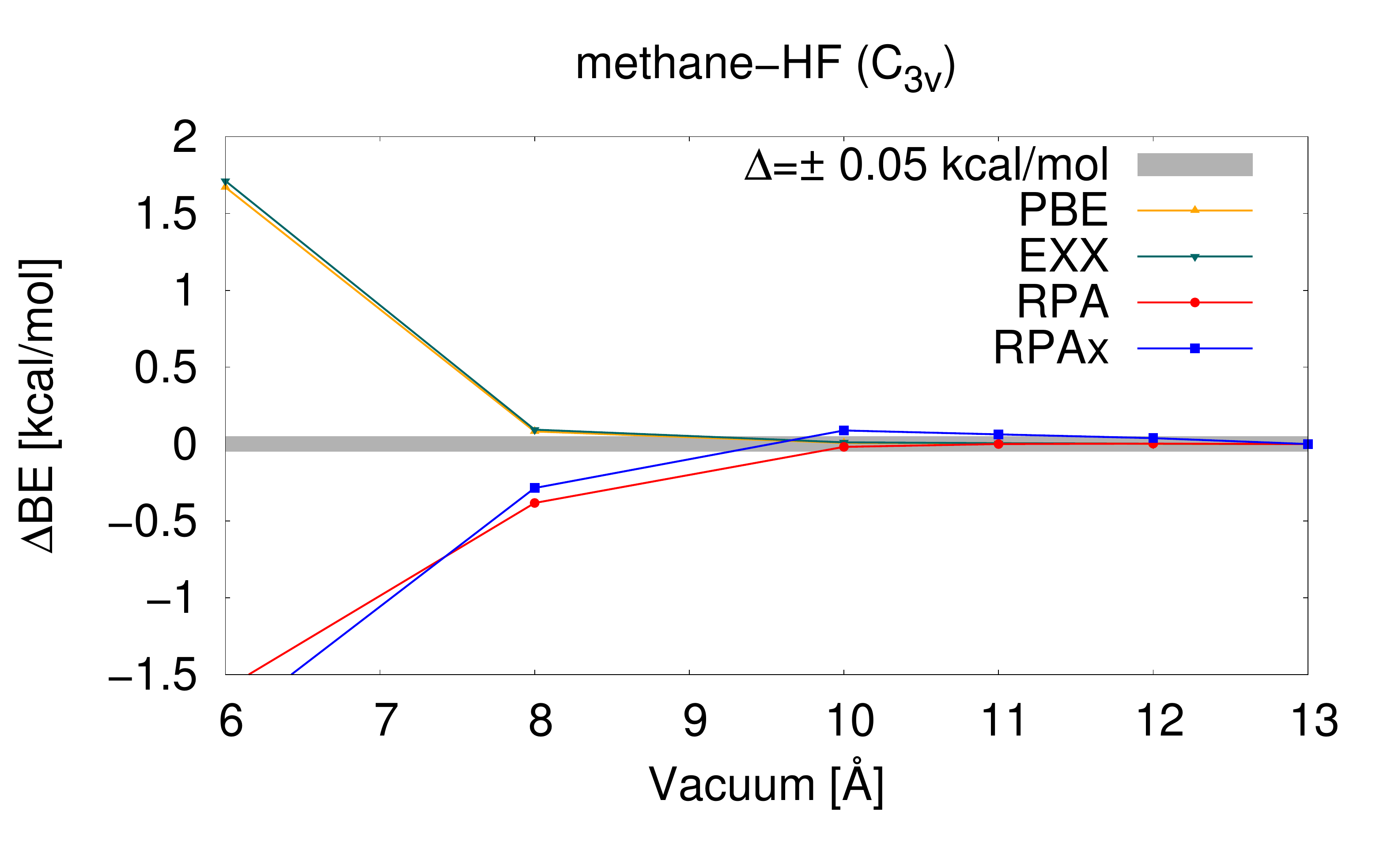}
 \includegraphics[height=0.2\textwidth]{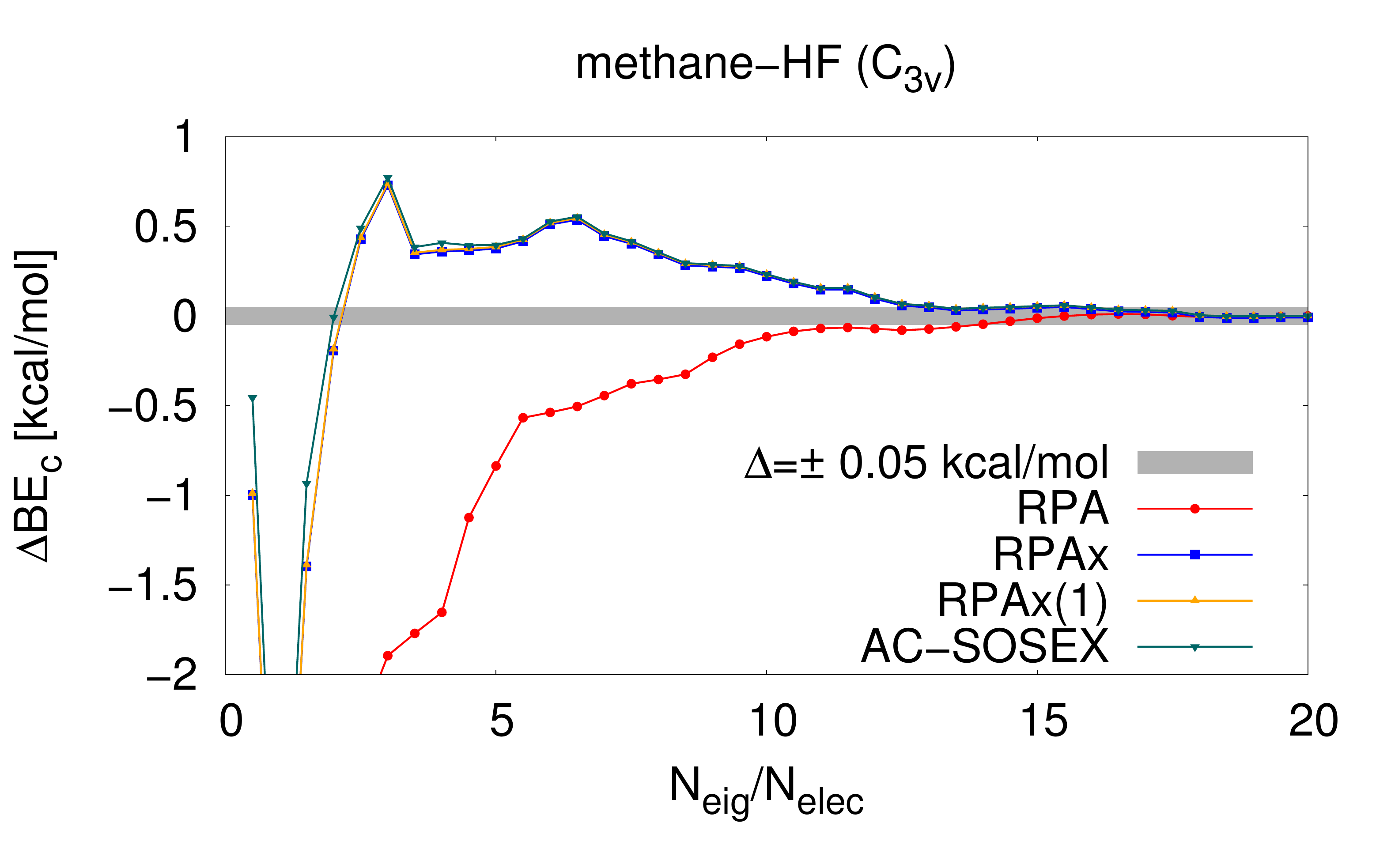}
 \includegraphics[height=0.2\textwidth]{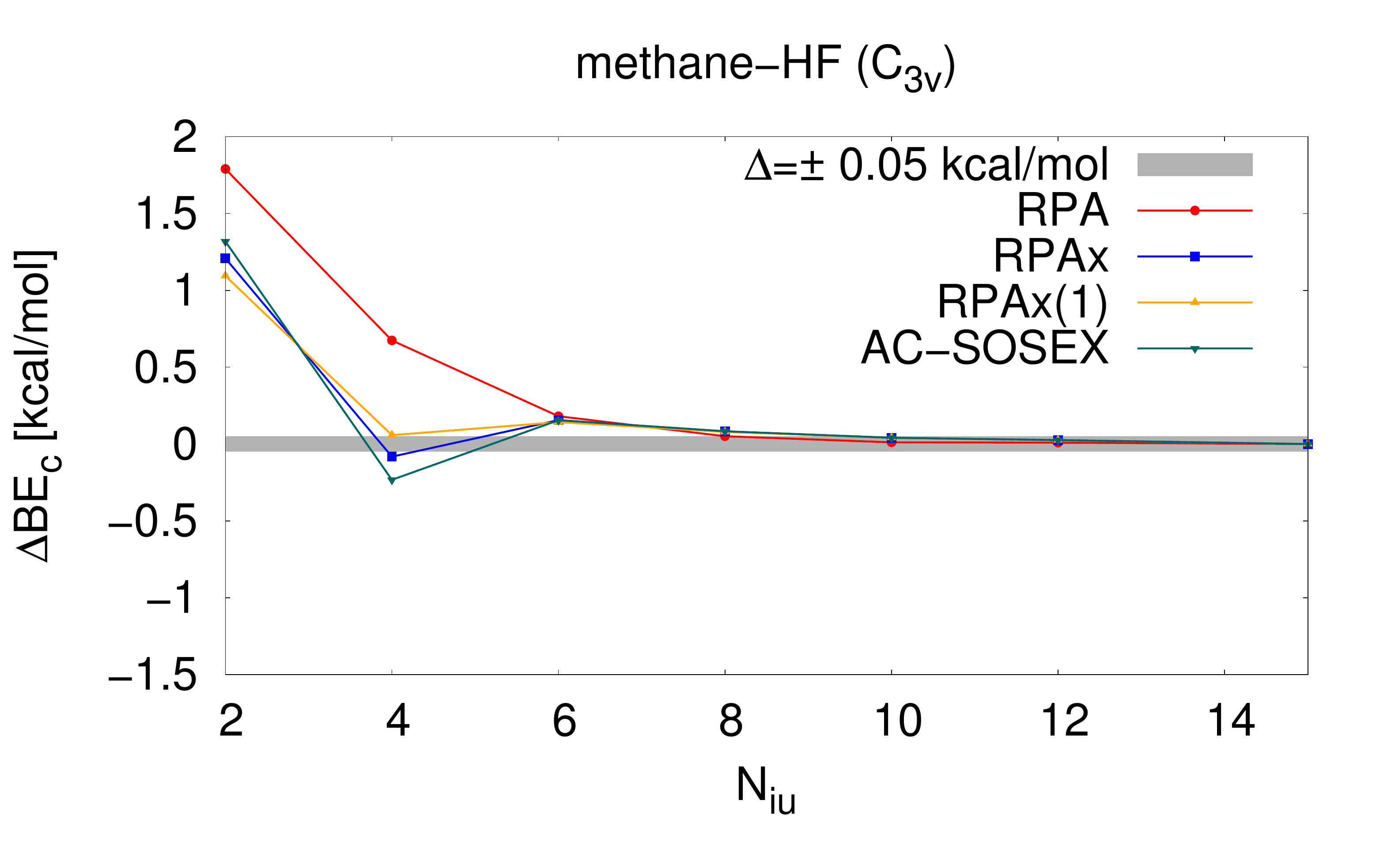}
 \includegraphics[height=0.2\textwidth]{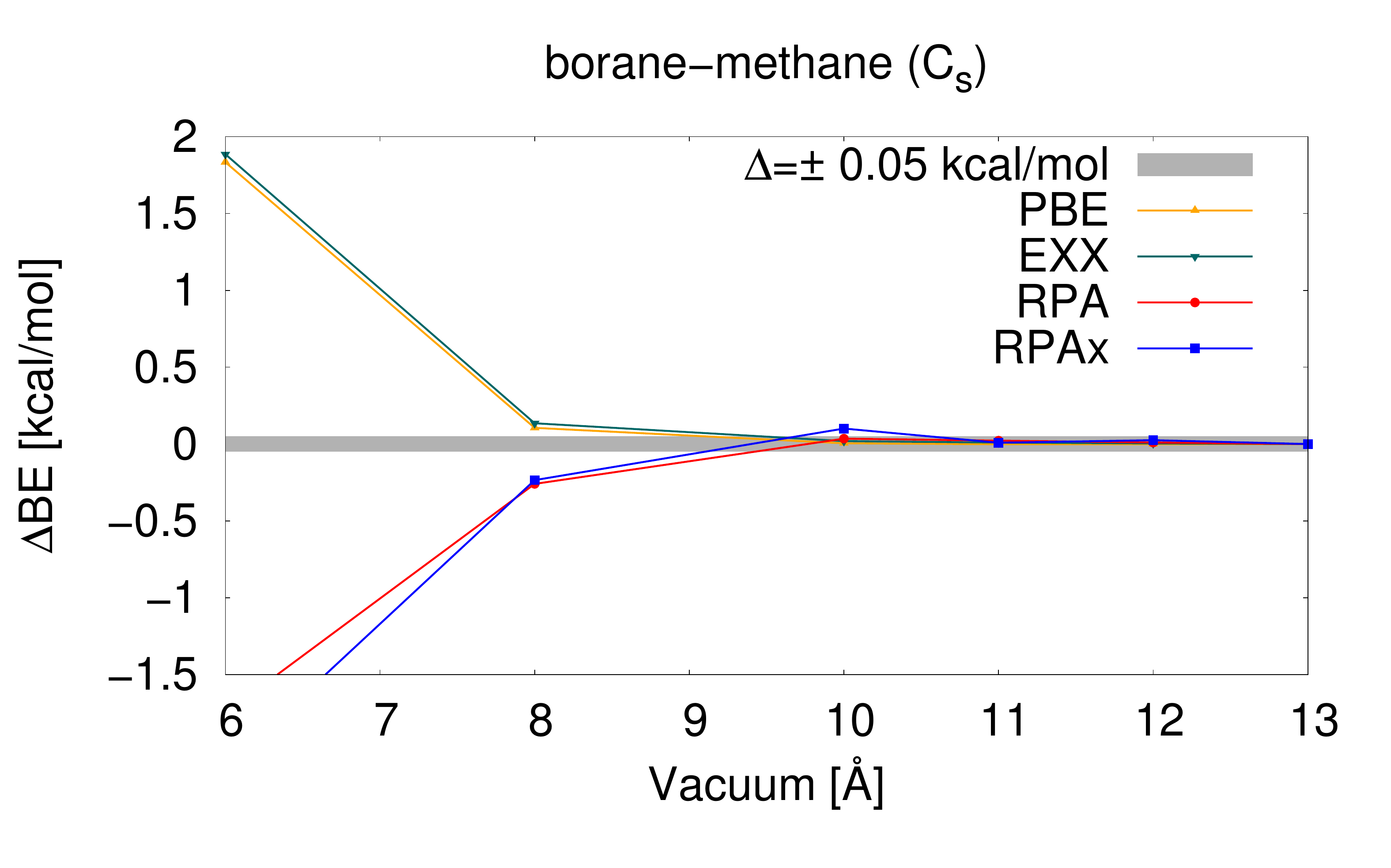}
 \includegraphics[height=0.2\textwidth]{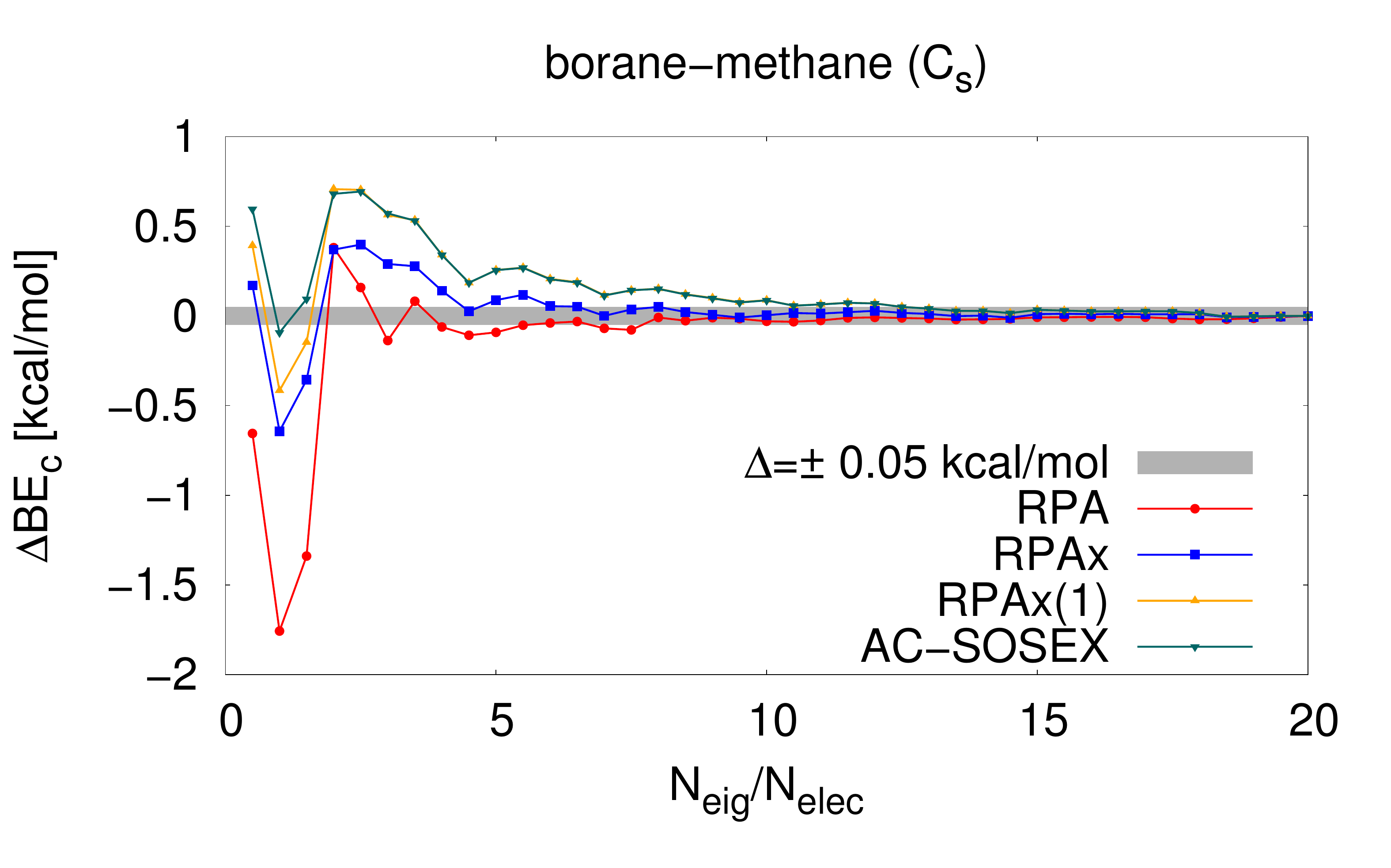}
 \includegraphics[height=0.2\textwidth]{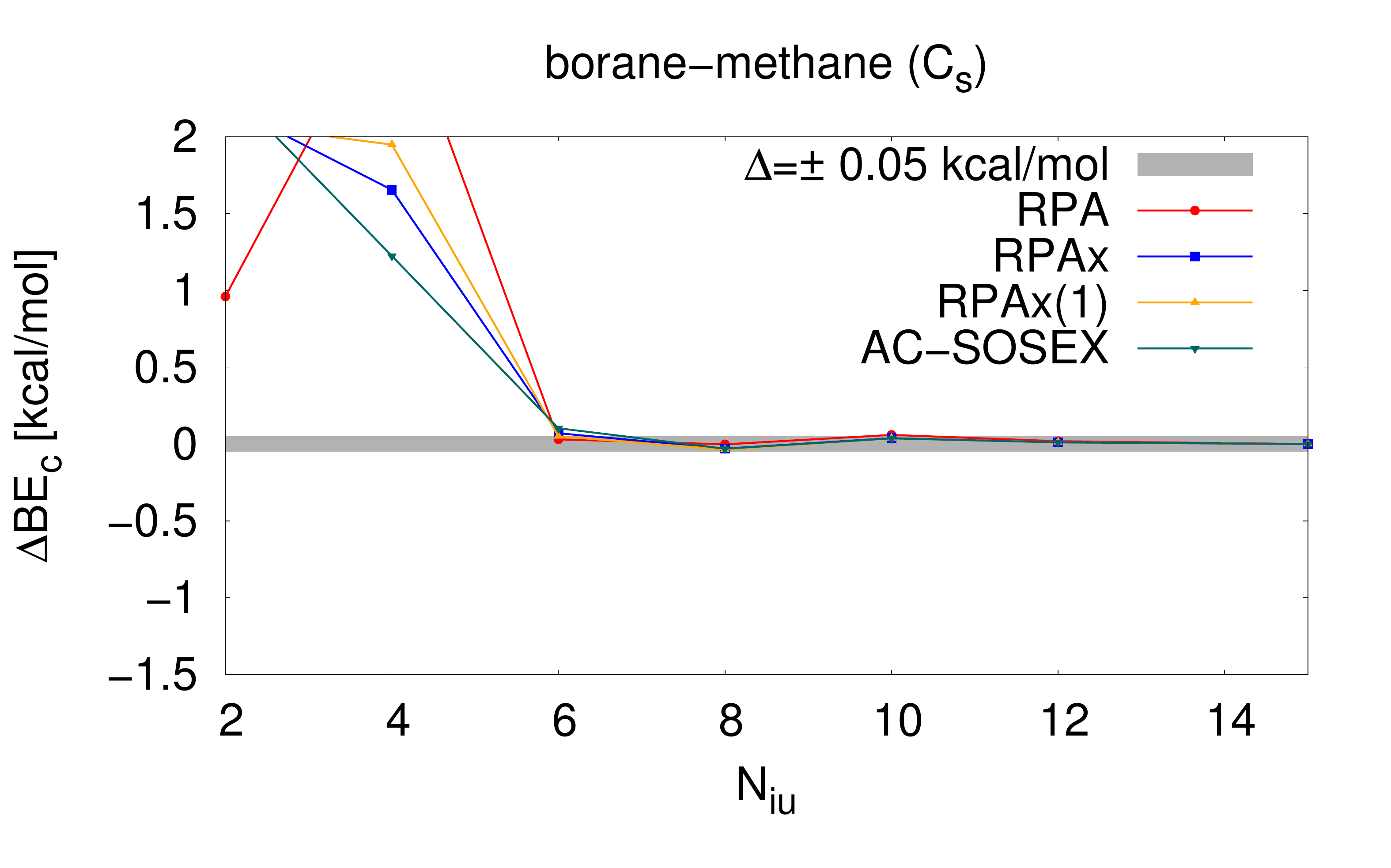}
 \caption{Convergence of the binding energy at all the level of the theory presented in the text with respect to vacuum (left panel), number of eigenvalues $N_{\rm eig}$ (central panel) and number of frequencies $N_{\rm iu}$ (right panel) for three representative molecules in the A24 test-set. In each plot and
for each approximation, the zero has been set by the most converged calculation.}
 \label{fig:conv}
\end{figure*}

All the ACFDT calculations have been performed in a post-DFT fashion. 
The KS single-particle wavefunctions and energies needed as input for the 
ACFDT total energy calculation have been calculated at the PBE~\cite{pbe} 
level using an energy cut-off of 80 Ry; Optimized Norm-Conserving Vanderbilt
(ONCV) pseudopotentials~\cite{hamann_optimized_2013, schlipf_optimization_2015, ONCV_website}  
have been used to model the electron-ion interaction. The molecules have been placed in an orthorhombic 
cell with 12 $A$ of vacuum in each direction, sufficient to suppress the spurious 
interactions between periodic replica and to converge all the components of the binding
energy within 0.05 kcal/mol, as illustrated in the left panel of Fig.~\ref{fig:conv}. 

The exact xc-energy [second term on the rhs of Eq~(\ref{echi})] can be further separated into the KS exact-exchange (EXX) energy 
\be
E_{\rm x} =-\frac{1}{2}\int\! d{\bf r} d{\bf r}' \frac{|\sum_i^{\rm occ}\phi_i({\bf r})\phi^*_i({\bf r}')|^2}{|{\bf r}-{\bf r}'|}
  \label{eq:exx}
\ee
and the ACFDT correlation energy determined by the difference between the interacting and 
non-interacting KS response functions
\be
E_{\rm c} =-\int_0^1 d\lambda \int_0^{\infty} \frac{du}{2\pi}\; \Tr \left\{ v \left[ \chi_{\lambda}(iu) 
 - \chi_s(iu) \right] \right\}.
  \label{eq:Ec_acfdt}
\ee
The integrable divergence appearing in a plane-wave implementation of the EXX energy would lead to a slow convergence with respect to the size of the supercell. In this work this issue
has been dealt with using the method proposed by Gygi and Baldereschi~\cite{gygi_self-consistent_1986} plus the extrapolation scheme of Nguyen and de Gironcoli~\cite{nguyen_efficient_2009}, although 
other strategies are also possible.~\cite{spencer_efficient_2008,marsili_method_2013} 
For all cases analyzed, this correction scheme allows the EXX contribution to converge 
the Binding Energy (BE) within 0.05 kcal/mol with 10 $A$ of vacuum separating the molecular replicas.

The strategy we use to solve the Dyson equation for $\chi_\l$ and efficiently compute the trace over the spatial coordinates in Eq.~(\ref{eq:Ec_acfdt}) is based on the solution of a well defined generalized eigenvalue 
problem (GEP) involving the non-interacting response function $\chi_s$ and its first 
order correction in the limit of vanishing coupling-constant $\mathcal{H}_{\rm Hx} = 
\chi_s (v+f_{\rm x}) \chi_s$. 
The set of eigenvectors of the GEP defines an optimal basis set on which i) the 
response matrices have a compact representation and ii) the Dyson equation has a 
straightforward solution, as detailed in Appendix~\ref{sec:appendix}.
Most importantly, only the lowest-lying eigenvectors and eigenvalues of the GEP at hand are relevant 
for the correlation energy calculation, and an efficient iterative diagonalization techniques 
can be used to compute the first $N_{\rm eig}$ lowest-lying eigenvalues/eigenvectors of the problem.
Moreover, the matrix elements of $\chi_s$ and of $\mathcal{H}_{\rm x} = \chi_s f_{\rm x} \chi_s$, 
needed for the solution of the GEPs, can be efficiently computed~\cite{nguyen_efficient_2009,PhysRevB.90.125150}
resorting to the linear response machinery of density-functional perturbation theory~\cite{baroni_phonons_2001} (DFPT)
thus avoiding explicit summations over empty states.

Once the GEP is solved, the trace over spatial coordinates can be written as a sum over a 
simple function $\mathcal{E}_c$ of the GEP eigenvalue $e_{\a}$ (and eventually of the diagonal
matrix elements of $\chi_s$ and $\mathcal{H}_{\rm x}$; 
see Appendix~\ref{sec:appendix} for the actual expression of $\mathcal{E}_c$), and 
a compact expression for the correlation energy is obtained:
\be
E_{\rm c} = -\int_0^1 d\lambda \int_0^{\infty} \frac{du}{2\pi}\; \sum_{\a=1}^{N_{\rm eig}} \mathcal{E}_c[e_{\a}(iu,\l),\l].
\label{eq:ec_acfdt_eig}
\ee
Here the frequency integral has been moved from the real to the imaginary axis of the complex plain~\footnote{
This can be done since that the density–density response functions are analytic in the upper
half-plane and vanish at infinity faster than $\omega^{-1}$ (see for instance Ref.~\onlinecite{giuliani_quantum_2005}).
}
to take advantage of the smooth behavior of the response functions for an efficient numerical quadrature (see below).
The integration over the coupling constant is analytic for all the approximations but the RPAx(1)
for which the eigenvalues $e^{\rm RPAx(1)}_{\a}$ inherit a non trivial dependence on the coupling constant 
from the $\lambda$-dependent RPAx(1) GEP (see Appendix~\ref{sec:appendix_rpax1}).
In this case a numerical integration can be efficiently performed using a Gauss-Legendre quadrature.
Usually very few points are needed to converge the integral within 0.01 kcal/mol (see Fig.~\ref{fig:conv_lambda}). 
For all the other approximations the GEPs do not depend on $\l$ and 
only the explicit dependence of $\mathcal{E}_c$ on the coupling constant has to
be considered. In these cases a straightforward analytic integration is always possible.

The number of eigenvalues N$_{eig}$ to compute is a convergence parameter. The central panel of 
Fig.~\ref{fig:conv} shows the dependence of the correlation binding energy on N$_{eig}$ for the
selected subset of molecules. In all the cases and for all the approximations studied, a relatively
small number of eigenvalues, never larger than 20 times the number of electrons in the system at hand,
is sufficient to converge the ACFDT correlation energy contribution to the BE within 0.05 kcal/mol.  

Finally, the integral over the imaginary frequency $iu$ can be efficiently performed using a Gauss-Legendre 
quadrature. A standard mesh of point $z_i \in [0,\pi/2]$ is mapped on an imaginary-frequency 
grid between $[0,+\infty)$ using the transformation $u_i = (\varepsilon_{HL} -cz_i)\tan(z_i)$ where $\varepsilon_{HL}$ 
is the DFT HOMO-LUMO gap of the systems and the parameter $c$ is set by the upper integration limit ($u_{\rm max}=200$ Ry).
This transformation takes into account the typical dependence of the response functions on the imaginary frequency
and the fact that at small frequency the scale of the excitation energies is given by the HOMO-LUMO gap. 
With this strategy a grid with $N_{\rm iu}=10$ points is usually sufficient to converge the 
ACFDT energy within 0.05 kcal/mol as illustrated in the right panel of Fig.~\ref{fig:conv}.

\section{Conclusions}
In this work we have defined a local EXX vertex based on the 
EXX kernel of TDDFT. This allowed us to unify 
different beyond-RPA approaches such as the various re-summations 
of RPAx and the SOSEX approximation, as well as to give a many-body 
perspective on approximations normally defined within DFT.

We have tested the theory on the H$_2$ molecule and the electron gas and we find that our AC-SOSEX agrees well with previous results in the literature. 
Although AC-SOSEX gives excellent total energies for the electron gas, 
it cannot be well represented as an approximation to the density 
response function. This could explain the poor performance in the dissociation 
region of H$_2$. On the contrary, approximations that incorporate the vertex 
correction in both the screened interaction and in the self-energy (RPA, RPAx and RPAx(1)) dissociate H$_2$ correctly. 

In terms of an overall performance we find that the RPAx(1) gives the most 
reliable results. It is as accurate as RPAx but shows no pathologies for 
any of the systems studied so far. High quality results within a reasonable computational cost are obtained on the A24 test-set, paving the way for accurate {\em ab-initio} treatment of sparse systems.
\begin{table*}[t]
\caption{Binding energies (in kcal/mol) for the A24 test-set as obtained with the RPA, RPAx, RPAx(1) and AC-SOSEX methods Compared with CCSDT(Q) reference values~\cite{rezac_describing_2013}}
\begin{ruledtabular}
\begin{tabular}{l l l r p{1cm} c c c c c }
    &  & Symm & RPA & RPAx & RPAx(1) & AC-SOSEX & CCSDT(Q) \\
\hline
   & {\it Hydrogen-bonded systems} & & & &  & \\ 
01 &     water-ammonia      & $C_s$     &  -5.71  &  -6.20  &  -6.14    &  -6.51  &  -6.492  \\
02 &     water dimer        & $C_s$     &  -4.27  &  -4.89  &  -4.82    &  -5.16  &  -4.994  \\
03 &     HCN dimer          & $C_s$     &  -4.20  &  -4.72  &  -4.50    &  -4.94  &  -4.738  \\
04 &     HF dimer           & $C_s$     &  -3.86  &  -4.47  &  -4.42    &  -4.78  &  -4.564  \\
05 &     ammonia dimer      & $C_{2h}$  &  -2.65  &  -2.92  &  -3.00    &  -3.12  &  -3.141  \\
\noalign{\vskip 2mm}  
   &     ME                 &           &   0.64  &   0.14  &   0.21    &  -0.12  & --- \\ 
   &     MAE                &           &   0.64  &   0.14  &   0.21    &   0.13  & --- \\
   &    MA\%E               &           &   13.8\%  & 3.2\% &   4.3\%   &   2.7\%  & --- \\
\noalign{\vskip 1mm}
\hline                   
\noalign{\vskip 1mm}
   & {\it Mixed-type systems} & & & &  & \\
06 &     methane-HF 	    & $C_{3v}$  &  -1.33   &  -1.70  &  -1.71   &  -1.91  &  -1.660  \\
07 &     ammonia-methane    & $C_{3v}$  &  -0.54   &  -0.86  &  -0.86    &  -0.92  &  -0.771  \\
08 &     methane-water 	    & $C_s$     &  -0.54   &  -0.76  &  -0.76   &  -0.82  &  -0.665  \\
09 &     formaldehyde dimer & $C_s$     &  -2.78   &  -3.82  &  -4.33   &  -5.23  &  -4.479  \\
10 &      ethene-water 	    & $C_s$     &  -2.11   &  -2.57  &  -2.59   &  -2.92  &  -2.564  \\
11 &    ethene-formaldehyde & $C_s$     &  -1.32   &  -1.78  &  -1.73   &  -1.94  &  -1.623  \\
12 &     ethyne dimer 	    & $C_{2v}$  &  -1.28   &  -1.66  &  -1.59   &  -1.79  &  -1.529  \\
13 &     ethene-ammonia     & $C_s$     &  -1.05   &  -1.43  &  -1.44   &  -1.58  &  -1.382  \\
14 &     ethene-dimer 	    & $C_{2v}$  &  -0.71   &  -1.13  &  -1.15   &  -1.24  &  -1.106  \\
15 &     methane-ethene     & $C_s$     &  -0.34   &  -0.64  &  -0.61   &  -0.66  &  -0.509  \\
\noalign{\vskip 2mm}
   &     ME                 &           &  0.43   &  -0.01  & -0.05    & -0.27  & --- \\
   &     MAE                &           &  0.43   &  0.14   &  0.08    &  0.27  & --- \\
   &    MA\%E               &           &  25.2\%   &  9.3\%   &  7.2\%    &  18\%  & --- \\
\noalign{\vskip 1mm}
\hline
\noalign{\vskip 1mm}
   & {\it Dispersion-dominated bonds} & & & & & & \\
16 &    borane-methane 	    & $C_s$     &  -0.57   &  -1.23  &  -1.39   & -1.58   & -1.513  \\
17 &    methane-ethane 	    & $C_s$     &  -0.58   &  -1.02  &  -1.00   & -1.03   & -0.836  \\
18 &    methane-ethane      & $C_3$     &  -0.44   &  -0.78  &  -0.75   & -0.78   & -0.614  \\
19 &    methane-dimer  	    & $C_{3d}$  &  -0.38   &  -0.53  &  -0.52   &  -0.52  & -0.539  \\
20 &    methane-Ar  	    & $C_{3v}$  &  -0.32   &  -0.42  &  -0.41   &  -0.42  & -0.405  \\
21 &    ethene-Ar   	    & $C_{2v}$  &  -0.27   &  -0.35  &  -0.35   &  -0.35  & -0.365  \\
22 &    ethene-ethyne 	    & $C_{2v}$  &  +1.00   &  +1.04  &  +0.96   &  +1.15  & +0.794  \\
23 &    ethene dimer 	    & $D_{2h}$  &  +1.11   &  +1.18  &  +1.09   &  +1.35  & +0.909  \\
24 &    ethyne dimer 	    & $D_{2h}$  &  +1.30   &  +1.30  &  +1.20   &  +1.38  & +1.084  \\
\noalign{\vskip 2mm}
   &     ME                 &           &  0.26     &  0.08  & 0.03     &  0.08  & --- \\
   &     MAE                &           &  0.26     & 0.15   & 0.10     &  0.17  & --- \\
   &   MA\%E                &           &   29.5\%    &  17.4\%  & 12.3\%    & 20.1\%   & --- \\
\noalign{\vskip 1mm}
 \end{tabular}
\end{ruledtabular}
\end {table*}
\appendix 

\section{The auxiliary basis sets}
\label{sec:appendix}

We provide in this section additional information on how the auxiliary basis set, used 
to solve the Dyson equation and calculate the trace over spatial coordinates, is computed 
for each approximation described in the text. 

\subsection{The RPA correlation energy}
\label{sec:appendix_rpa}
The evaluation of the RPA correlation energy is based on an eigenvalue decomposition of the noninteracting
response function and has been carefully detailed in Ref.~\onlinecite{nguyen_efficient_2009}; we summarize here 
the main features of the RPA implementation because the solution of the 
RPA problem is a prerequisite for all the exchange-corrected approximations. 

For each point on the imaginary-frequency grid the generalized eigenvalue problem (GEP) 
\be
v \chi_s |w_i\rangle = e_i |w_i\rangle,
\label{eq:GEP_rpa}
\ee
is solved (the dependence on the imaginary frequency is implicitly assumed). Once the solution of the GEP is available
the RPA Dyson equation, Eq.~(\ref{eq:dyson_rpa}), can be readily solved since $\{w_i\}$ are also eigenvectors
of $v\chi_{\lambda}^{\rm RPA}$ with eigenvalues $e^{\rm RPA}_i = e_i/(1-\lambda e_i)$. The trace 
over spatial coordinates appearing in the expression for the RPA correlation energy can be written as 
\be
\Tr\{v[\chi^{\rm RPA}_{\lambda}(iu)-\chi_s(iu)] \} = \sum_i \left[ \frac{e_i(iu)}{1-\lambda e_i(iu)} - e_i(iu) \right].
\ee
The integration over the coupling constant can be performed analytically and the final 
result for the RPA correlation energy becomes:
\be
E_{\rm c}^{\rm RPA} = \frac{1}{2 \pi} \int_0^\infty du\; \sum_i \left[ \ln[1-e_i(iu)] + e_i(iu) \right].
\ee
The spectrum of the GEP in Eq.~(\ref{eq:GEP_rpa}) is bounded from above by zero, and only
a small number of the lowest lying eigenvalues are significantly different from zero.~\cite{baldereschi_dielectric_1979,wilson_efficient_2008}
This implies that only a small fraction of the $\{e_i\}$ contributes significantly to the correlation energy,
and an efficient iterative diagonalization scheme can be used to evaluate those.~\cite{nguyen_efficient_2009,viet_phd_thesis} The basic operation involved in the iterative solution of the 
GEP in Eq.~(\ref{eq:GEP_rpa}) is the calculation of the noninteracting response to a trial potential, and 
this is done resorting to the linear-response techniques of density functional perturbation theory~\cite{baroni_phonons_2001}
(DFPT), generalized to imaginary frequencies. 

\subsection{The RPAx correlation energy}
\label{sec:appendix_rpax}
A strategy analogous to the one adopted for the RPA problem is used to compute the RPAx correlation energy.
The GEP to solve in this case is the following
\be
-\chi_s\left[ v + f_{\rm x} \right]\chi_s |z^{\rm RPAx}_{\alpha}\rangle = e^{\rm RPAx}_{\alpha} [-\chi_s] |z^{\rm RPAx}_{\alpha} \rangle.
\label{eq:GEP_rpax}
\ee
The expression for $\mathcal{H}_{\rm x} = \chi_sf_{\rm x} \chi_s$ in terms of KS orbitals and eigenvalues 
has been obtained by G\"orling,~\cite{gorling_exact_1998_pra,gorling_exact_1998} and Hellgren and von Barth.~\cite{hvb08} 
In this case the diagonalization was carried out in the basis set of the 
eigenvectors of the RPA GEP, Eq.~(\ref{eq:GEP_rpa}). This was done in order to i) avoid possible instabilities that may occur in the inversion of the non-interacting response function
(i.e. the overlap matrix in the GEP) and ii) to speed up the calculations. 
On this basis set $\chi_s$ and $\mathcal{H}_{\rm H}=\chi_s v \chi_s$ 
are diagonal and their matrix elements are readily available in terms of the eigenvalues of the RPA GEP (Eq.~(\ref{eq:GEP_rpa})), i.e.
$\chi_s^{ij} = \langle w_i| \chi_s | w_j \rangle = \delta_{ij}e_i$ and $\mathcal{H}_{\rm H}^{ij} = \langle w_i| \chi_s v \chi_s | w_j \rangle = \delta_{ij}e^2_i$. 
The only additional operation required to solve the RPAx problem is the evaluation of 
the matrix elements $\mathcal{H}_{\rm x}^{ij}=\langle w_i| \chi_s f_{\rm x} \chi_s | w_j \rangle$. These can be efficiently 
computed using DFPT as detailed in Ref.~\onlinecite{PhysRevB.90.125150,NsC_phd_thesis}.
Once the solution of the GEP in Eq.~(\ref{eq:GEP_rpax}) is available the action of $v \chi_{\lambda}^{\rm RPAx}$ on 
the eigenvectors $\{z_{\alpha}^{\rm RPAx}\}$ can be explicitly expressed making use of the RPAx Dyson equation Eq.~(\ref{respx}), i.e. 
$v \chi_{\lambda}^{\rm RPAx}|z^{\rm RPAx}_{\alpha} \rangle=v \chi_s|z^{\rm RPAx}_{\alpha}\rangle/(1-\lambda e^{\rm RPAx}_{\alpha})$, and the trace over spatial coordinates in the RPAx correlation energy can be written as 
\begin{align}
& \Tr\{ v [\chi_{\lambda}^{\rm RPAx}(iu) -\chi_s(iu)] \} =  \nonumber \\ 
&  \quad = \sum_{\alpha} \left. \langle z^{\rm RPAx}_{\alpha} | \chi_s v \chi_s | z^{\rm RPAx}_{\alpha} \rangle \right._{iu} \left[ 1-\frac{1}{1-\lambda e^{\rm RPAx}_{\alpha}(iu)} \right]. 
\end{align}
The $\lambda$ integration is analytic and the final result for the RPAx correlation energy reads
\begin{align}
 E_{\rm c}^{\rm RPAx} = -\frac{1}{2 \pi} \int_0^\infty & du\; \sum_{\alpha} \frac{\left. \langle z^{\rm RPAx}_{\alpha} | \chi_s v \chi_s | z^{\rm RPAx}_{\alpha} \rangle \right._{iu}}{e_{\alpha}^{\rm RPAx}(iu)} \nonumber \\
& \times  \left[ \ln[1-e^{\rm RPAx}_{\alpha}(iu)] + e^{\rm RPAx}_{\alpha}(iu) \right]
\end{align}
\subsection{The RPAx(1) correlation energy}
\label{sec:appendix_rpax1}
\begin{figure}[b]
 \includegraphics[height=0.3\textwidth]{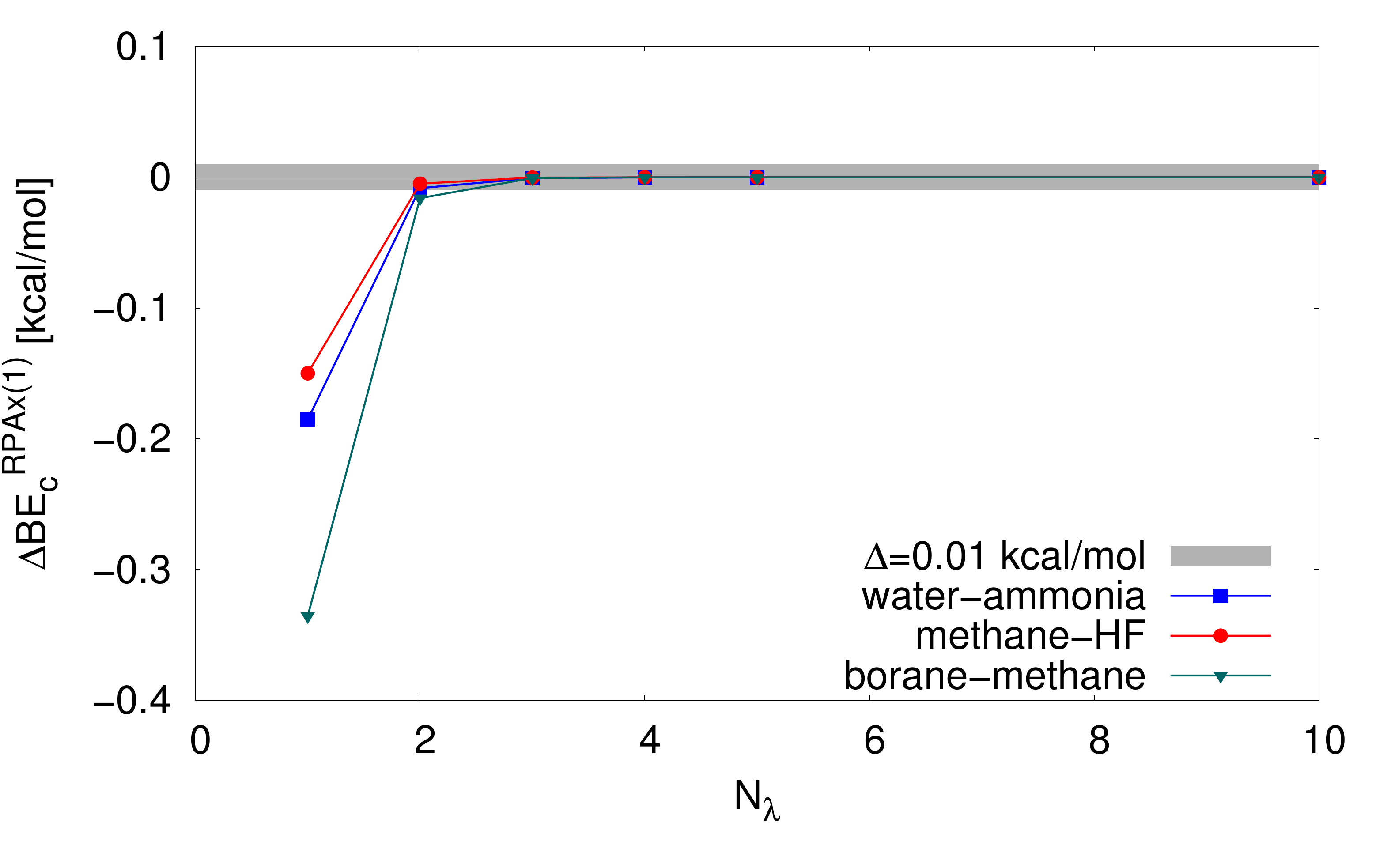}
 \caption{Convergence of the RPAx(1) binding energy with respect to the number of points in the $\l$ grid for three different 
 complexes.}
 \label{fig:conv_lambda}
\end{figure}
The RPAx(1) GEP is the following 
\be
v P_{\lambda}^{(1)} | z_{\l,\a}^{\rm RPAx(1)} \rangle = e^{\rm RPAx(1)}_{\l,\a}|z_{\l,\a}^{\rm RPAx(1)} \rangle
\label{eq:GEP_rpax1}
\ee
and it is solved in the basis set of the RPA eigenvectors [Eq.~(\ref{eq:GEP_rpa})]. 
Also in this case the only additional information needed is the representation of the $\mathcal{H}_{\rm x}$ operator on 
this basis (this may be already available if an RPAx calculation has been previously performed). The solution of the
RPAx(1) GEP together with the RPAx(1) Dyson equation [Eq.~(\ref{eq:dyson_rpax1})] allows to write explicitly 
$v\chi_{\lambda}^{\rm RPAx(1)} |z_{\a}^{\rm RPAx(1)} \rangle = e^{\rm RPAx(1)}_{\l,\a}|z_{\a}^{\rm RPAx(1)} \rangle 
/ (1-\l e^{\rm RPAx(1)}_{\l,\a})$, and the trace appearing in the correlation energy expression becomes
\begin{align}
&\Tr\{ v [\chi_{\lambda}^{\rm RPAx(1)}(iu) -\chi_s(iu)] \} =  \nonumber \\ 
& \quad = \sum_{\alpha} \frac{e_{\l,\alpha}^{\rm RPAx(1)}(iu)}{1-\lambda e^{\rm RPAx(1)}_{\l,\alpha}(iu)}
- \left. \langle z^{\rm RPAx(1)}_{\l,\alpha} | \chi_s | z^{\rm RPAx(1)}_{\l,\alpha} \rangle \right._{iu} 
\end{align}
At variance with previous cases, for the RPAx(1) problem the integration over the coupling 
constant has to be performed numerically because of the non-trivial dependence of the eigenvalues and
eigenvectors on coupling constant inherited from the $\l$-dependent GEP Eq.~(\ref{eq:GEP_rpax1}). However the 
integrand is a smooth function and a Gauss-Legendre quadrature with less than five points 
ensures converged results within 0.01 kcal/mol (see Fig.~\ref{fig:conv_lambda}). We stress here that the most 
computationally intensive part of the whole procedure is the evaluation of the response functions 
matrix elements. Once $\chi_s(iu)$ and $\mathcal{H}_{\rm Hx}(iu)$ have been represented, only linear algebra 
operations with matrices of dimension $N_{eig}\times N_{eig}$ are needed to solve the GEP for each 
$\l$ on the grids, meaning that the additional cost associated to the numerical coupling-constant
integration is only a tiny fraction of the total computational cost.
\subsection{The AC-SOSEX correlation energy}
\label{sec:appendix_sosex}
For the AC-SOSEX approximation the RPA eigenvectors $\{w_i\}$ from Eq.~(\ref{eq:dyson_sosex}) can be used to 
explicitly solve the SOSEX dyson equation [Eq.~(\ref{eq:dyson_sosex})] and compute the action of 
the SOSEX response function
\be
v\chi_{\l}^{\rm SOSEX} |w_i \rangle = \frac{(e_i + \l \mathcal{H}_{\rm x})|w_i\rangle}{1-\l e_i}.
\ee
The trace in the correlation energy expression can then be readily computed in the $\{|\w_i\rangle\}$ basis set:
\begin{align}
&\Tr\{ v [\chi_{\lambda}^{\rm SOSEX}(iu) -\chi_s(iu)] \} =  \nonumber \\ 
& \quad = \sum_i \frac{e_i(iu)+\l \mathcal{H}_{\rm x}^{ii}(iu)}{1-\lambda e_i(iu)} -e_i(iu).
\end{align}
The integration over the coupling constant is analytic and the final expression for the
correlation energy reads
\begin{align}
E_c^{\rm {AC-SOSEX}} = \frac{1}{2 \pi} & \int_0^\infty du\; \sum_i \frac{e^2_i(iu) -\mathcal{H}_{\rm x}^{ii}(iu)}{e_i^2} \nonumber \\
&  \times \left[ \ln (1-e_i(iu))+e_i \right]
\end{align}

%


\begin{thebibliography}{70}%
\makeatletter
\providecommand \@ifxundefined [1]{%
 \@ifx{#1\undefined}
}%
\providecommand \@ifnum [1]{%
 \ifnum #1\expandafter \@firstoftwo
 \else \expandafter \@secondoftwo
 \fi
}%
\providecommand \@ifx [1]{%
 \ifx #1\expandafter \@firstoftwo
 \else \expandafter \@secondoftwo
 \fi
}%
\providecommand \natexlab [1]{#1}%
\providecommand \enquote  [1]{``#1''}%
\providecommand \bibnamefont  [1]{#1}%
\providecommand \bibfnamefont [1]{#1}%
\providecommand \citenamefont [1]{#1}%
\providecommand \href@noop [0]{\@secondoftwo}%
\providecommand \href [0]{\begingroup \@sanitize@url \@href}%
\providecommand \@href[1]{\@@startlink{#1}\@@href}%
\providecommand \@@href[1]{\endgroup#1\@@endlink}%
\providecommand \@sanitize@url [0]{\catcode `\\12\catcode `\$12\catcode
  `\&12\catcode `\#12\catcode `\^12\catcode `\_12\catcode `\%12\relax}%
\providecommand \@@startlink[1]{}%
\providecommand \@@endlink[0]{}%
\providecommand \url  [0]{\begingroup\@sanitize@url \@url }%
\providecommand \@url [1]{\endgroup\@href {#1}{\urlprefix }}%
\providecommand \urlprefix  [0]{URL }%
\providecommand \Eprint [0]{\href }%
\providecommand \doibase [0]{http://dx.doi.org/}%
\providecommand \selectlanguage [0]{\@gobble}%
\providecommand \bibinfo  [0]{\@secondoftwo}%
\providecommand \bibfield  [0]{\@secondoftwo}%
\providecommand \translation [1]{[#1]}%
\providecommand \BibitemOpen [0]{}%
\providecommand \bibitemStop [0]{}%
\providecommand \bibitemNoStop [0]{.\EOS\space}%
\providecommand \EOS [0]{\spacefactor3000\relax}%
\providecommand \BibitemShut  [1]{\csname bibitem#1\endcsname}%
\let\auto@bib@innerbib\@empty
\bibitem [{\citenamefont {Hohenberg}\ and\ \citenamefont
  {Kohn}(1964)}]{hohenbergkohn}%
  \BibitemOpen
  \bibfield  {author} {\bibinfo {author} {\bibfnamefont {P.}~\bibnamefont
  {Hohenberg}}\ and\ \bibinfo {author} {\bibfnamefont {W.}~\bibnamefont
  {Kohn}},\ }\href {\doibase 10.1103/PhysRev.136.B864} {\bibfield  {journal}
  {\bibinfo  {journal} {Phys. Rev.}\ }\textbf {\bibinfo {volume} {136}},\
  \bibinfo {pages} {B864} (\bibinfo {year} {1964})}\BibitemShut {NoStop}%
\bibitem [{\citenamefont {Kohn}\ and\ \citenamefont
  {Sham}(1965)}]{kohnsham1965}%
  \BibitemOpen
  \bibfield  {author} {\bibinfo {author} {\bibfnamefont {W.}~\bibnamefont
  {Kohn}}\ and\ \bibinfo {author} {\bibfnamefont {L.~J.}\ \bibnamefont
  {Sham}},\ }\href {\doibase 10.1103/PhysRev.140.A1133} {\bibfield  {journal}
  {\bibinfo  {journal} {Phys. Rev.}\ }\textbf {\bibinfo {volume} {140}},\
  \bibinfo {pages} {A1133} (\bibinfo {year} {1965})}\BibitemShut {NoStop}%
\bibitem [{\citenamefont {Hay}\ \emph {et~al.}(2015)\citenamefont {Hay},
  \citenamefont {Ferlat}, \citenamefont {Casula}, \citenamefont {Seitsonen},\
  and\ \citenamefont {Mauri}}]{henri2015}%
  \BibitemOpen
  \bibfield  {author} {\bibinfo {author} {\bibfnamefont {H.}~\bibnamefont
  {Hay}}, \bibinfo {author} {\bibfnamefont {G.}~\bibnamefont {Ferlat}},
  \bibinfo {author} {\bibfnamefont {M.}~\bibnamefont {Casula}}, \bibinfo
  {author} {\bibfnamefont {A.~P.}\ \bibnamefont {Seitsonen}}, \ and\ \bibinfo
  {author} {\bibfnamefont {F.}~\bibnamefont {Mauri}},\ }\href {\doibase
  10.1103/PhysRevB.92.144111} {\bibfield  {journal} {\bibinfo  {journal} {Phys.
  Rev. B}\ }\textbf {\bibinfo {volume} {92}},\ \bibinfo {pages} {144111}
  (\bibinfo {year} {2015})}\BibitemShut {NoStop}%
\bibitem [{\citenamefont {Roman-Roman}\ and\ \citenamefont
  {Zicovich-Wilson}(2015)}]{zeolites}%
  \BibitemOpen
  \bibfield  {author} {\bibinfo {author} {\bibfnamefont {E.~I.}\ \bibnamefont
  {Roman-Roman}}\ and\ \bibinfo {author} {\bibfnamefont {C.~M.}\ \bibnamefont
  {Zicovich-Wilson}},\ }\href {\doibase
  https://doi.org/10.1016/j.cplett.2014.11.044} {\bibfield  {journal} {\bibinfo
   {journal} {Chemical Physics Letters}\ }\textbf {\bibinfo {volume} {619}},\
  \bibinfo {pages} {109 } (\bibinfo {year} {2015})}\BibitemShut {NoStop}%
\bibitem [{\citenamefont {Varsano}\ \emph {et~al.}(2017)\citenamefont
  {Varsano}, \citenamefont {Sorella}, \citenamefont {Sangalli}, \citenamefont
  {Barborini}, \citenamefont {Corni}, \citenamefont {Molinari},\ and\
  \citenamefont {Rontani}}]{nanoexc}%
  \BibitemOpen
  \bibfield  {author} {\bibinfo {author} {\bibfnamefont {D.}~\bibnamefont
  {Varsano}}, \bibinfo {author} {\bibfnamefont {S.}~\bibnamefont {Sorella}},
  \bibinfo {author} {\bibfnamefont {D.}~\bibnamefont {Sangalli}}, \bibinfo
  {author} {\bibfnamefont {M.}~\bibnamefont {Barborini}}, \bibinfo {author}
  {\bibfnamefont {S.}~\bibnamefont {Corni}}, \bibinfo {author} {\bibfnamefont
  {E.}~\bibnamefont {Molinari}}, \ and\ \bibinfo {author} {\bibfnamefont
  {M.}~\bibnamefont {Rontani}},\ }\href {\doibase 10.1038/s41467-017-01660-8}
  {\bibfield  {journal} {\bibinfo  {journal} {Nat. Comm.}\ }\textbf {\bibinfo
  {volume} {8}},\ \bibinfo {pages} {1461} (\bibinfo {year} {2017})}\BibitemShut
  {NoStop}%
\bibitem [{\citenamefont {Heyd}\ \emph {et~al.}(2003)\citenamefont {Heyd},
  \citenamefont {Scuseria},\ and\ \citenamefont {Ernzerhof}}]{hse06}%
  \BibitemOpen
  \bibfield  {author} {\bibinfo {author} {\bibfnamefont {J.}~\bibnamefont
  {Heyd}}, \bibinfo {author} {\bibfnamefont {G.~E.}\ \bibnamefont {Scuseria}},
  \ and\ \bibinfo {author} {\bibfnamefont {M.}~\bibnamefont {Ernzerhof}},\
  }\href {\doibase 10.1063/1.1564060} {\bibfield  {journal} {\bibinfo
  {journal} {The Journal of Chemical Physics}\ }\textbf {\bibinfo {volume}
  {118}},\ \bibinfo {pages} {8207} (\bibinfo {year} {2003})}\BibitemShut
  {NoStop}%
\bibitem [{\citenamefont {Hedin}(1965)}]{hedin}%
  \BibitemOpen
  \bibfield  {author} {\bibinfo {author} {\bibfnamefont {L.}~\bibnamefont
  {Hedin}},\ }\href {\doibase 10.1103/PhysRev.139.A796} {\bibfield  {journal}
  {\bibinfo  {journal} {Phys. Rev.}\ }\textbf {\bibinfo {volume} {139}},\
  \bibinfo {pages} {A796} (\bibinfo {year} {1965})}\BibitemShut {NoStop}%
\bibitem [{\citenamefont {Dahlen}\ and\ \citenamefont {von
  Barth}(2004)}]{dahlenvar}%
  \BibitemOpen
  \bibfield  {author} {\bibinfo {author} {\bibfnamefont {N.~E.}\ \bibnamefont
  {Dahlen}}\ and\ \bibinfo {author} {\bibfnamefont {U.}~\bibnamefont {von
  Barth}},\ }\href {\doibase 10.1103/PhysRevB.69.195102} {\bibfield  {journal}
  {\bibinfo  {journal} {Phys. Rev. B}\ }\textbf {\bibinfo {volume} {69}},\
  \bibinfo {pages} {195102} (\bibinfo {year} {2004})}\BibitemShut {NoStop}%
\bibitem [{\citenamefont {Dahlen}\ \emph {et~al.}(2006)\citenamefont {Dahlen},
  \citenamefont {van Leeuwen},\ and\ \citenamefont {von
  Barth}}]{dahlenbarth06}%
  \BibitemOpen
  \bibfield  {author} {\bibinfo {author} {\bibfnamefont {N.~E.}\ \bibnamefont
  {Dahlen}}, \bibinfo {author} {\bibfnamefont {R.}~\bibnamefont {van Leeuwen}},
  \ and\ \bibinfo {author} {\bibfnamefont {U.}~\bibnamefont {von Barth}},\
  }\href {\doibase 10.1103/PhysRevA.73.012511} {\bibfield  {journal} {\bibinfo
  {journal} {Phys. Rev. A}\ }\textbf {\bibinfo {volume} {73}},\ \bibinfo
  {pages} {012511} (\bibinfo {year} {2006})}\BibitemShut {NoStop}%
\bibitem [{\citenamefont {Hellgren}\ and\ \citenamefont {von
  Barth}(2007)}]{hvb07}%
  \BibitemOpen
  \bibfield  {author} {\bibinfo {author} {\bibfnamefont {M.}~\bibnamefont
  {Hellgren}}\ and\ \bibinfo {author} {\bibfnamefont {U.}~\bibnamefont {von
  Barth}},\ }\href {\doibase 10.1103/PhysRevB.76.075107} {\bibfield  {journal}
  {\bibinfo  {journal} {Phys. Rev. B}\ }\textbf {\bibinfo {volume} {76}},\
  \bibinfo {pages} {075107} (\bibinfo {year} {2007})}\BibitemShut {NoStop}%
\bibitem [{\citenamefont {Freeman}(1977)}]{freeman}%
  \BibitemOpen
  \bibfield  {author} {\bibinfo {author} {\bibfnamefont {D.~L.}\ \bibnamefont
  {Freeman}},\ }\href@noop {} {\bibfield  {journal} {\bibinfo  {journal} {Phys.
  Rev. B}\ }\textbf {\bibinfo {volume} {15}},\ \bibinfo {pages} {5512}
  (\bibinfo {year} {1977})}\BibitemShut {NoStop}%
\bibitem [{\citenamefont {Gr\"uneis}\ \emph {et~al.}(2009)\citenamefont
  {Gr\"uneis}, \citenamefont {Marsman}, \citenamefont {Harl}, \citenamefont
  {Schimka},\ and\ \citenamefont {Kresse}}]{doi:10.1063/1.3250347}%
  \BibitemOpen
  \bibfield  {author} {\bibinfo {author} {\bibfnamefont {A.}~\bibnamefont
  {Gr\"uneis}}, \bibinfo {author} {\bibfnamefont {M.}~\bibnamefont {Marsman}},
  \bibinfo {author} {\bibfnamefont {J.}~\bibnamefont {Harl}}, \bibinfo {author}
  {\bibfnamefont {L.}~\bibnamefont {Schimka}}, \ and\ \bibinfo {author}
  {\bibfnamefont {G.}~\bibnamefont {Kresse}},\ }\href@noop {} {\bibfield
  {journal} {\bibinfo  {journal} {The Journal of Chemical Physics}\ }\textbf
  {\bibinfo {volume} {131}},\ \bibinfo {pages} {154115} (\bibinfo {year}
  {2009})}\BibitemShut {NoStop}%
\bibitem [{\citenamefont {Ren}\ \emph {et~al.}(2013)\citenamefont {Ren},
  \citenamefont {Rinke}, \citenamefont {Scuseria},\ and\ \citenamefont
  {Scheffler}}]{PhysRevB.88.035120}%
  \BibitemOpen
  \bibfield  {author} {\bibinfo {author} {\bibfnamefont {X.}~\bibnamefont
  {Ren}}, \bibinfo {author} {\bibfnamefont {P.}~\bibnamefont {Rinke}}, \bibinfo
  {author} {\bibfnamefont {G.~E.}\ \bibnamefont {Scuseria}}, \ and\ \bibinfo
  {author} {\bibfnamefont {M.}~\bibnamefont {Scheffler}},\ }\href {\doibase
  10.1103/PhysRevB.88.035120} {\bibfield  {journal} {\bibinfo  {journal} {Phys.
  Rev. B}\ }\textbf {\bibinfo {volume} {88}},\ \bibinfo {pages} {035120}
  (\bibinfo {year} {2013})}\BibitemShut {NoStop}%
\bibitem [{\citenamefont {Ren}\ \emph {et~al.}(2015)\citenamefont {Ren},
  \citenamefont {Marom}, \citenamefont {Caruso}, \citenamefont {Scheffler},\
  and\ \citenamefont {Rinke}}]{PhysRevB.92.081104}%
  \BibitemOpen
  \bibfield  {author} {\bibinfo {author} {\bibfnamefont {X.}~\bibnamefont
  {Ren}}, \bibinfo {author} {\bibfnamefont {N.}~\bibnamefont {Marom}}, \bibinfo
  {author} {\bibfnamefont {F.}~\bibnamefont {Caruso}}, \bibinfo {author}
  {\bibfnamefont {M.}~\bibnamefont {Scheffler}}, \ and\ \bibinfo {author}
  {\bibfnamefont {P.}~\bibnamefont {Rinke}},\ }\href {\doibase
  10.1103/PhysRevB.92.081104} {\bibfield  {journal} {\bibinfo  {journal} {Phys.
  Rev. B}\ }\textbf {\bibinfo {volume} {92}},\ \bibinfo {pages} {081104}
  (\bibinfo {year} {2015})}\BibitemShut {NoStop}%
\bibitem [{\citenamefont {Shirley}\ and\ \citenamefont
  {Martin}(1993)}]{PhysRevB.47.15404}%
  \BibitemOpen
  \bibfield  {author} {\bibinfo {author} {\bibfnamefont {E.~L.}\ \bibnamefont
  {Shirley}}\ and\ \bibinfo {author} {\bibfnamefont {R.~M.}\ \bibnamefont
  {Martin}},\ }\href {\doibase 10.1103/PhysRevB.47.15404} {\bibfield  {journal}
  {\bibinfo  {journal} {Phys. Rev. B}\ }\textbf {\bibinfo {volume} {47}},\
  \bibinfo {pages} {15404} (\bibinfo {year} {1993})}\BibitemShut {NoStop}%
\bibitem [{\citenamefont {Almbladh}\ \emph {et~al.}(1999)\citenamefont
  {Almbladh}, \citenamefont {von Barth},\ and\ \citenamefont {van
  Leeuwen}}]{almbladh}%
  \BibitemOpen
  \bibfield  {author} {\bibinfo {author} {\bibfnamefont {C.~O.}\ \bibnamefont
  {Almbladh}}, \bibinfo {author} {\bibfnamefont {U.}~\bibnamefont {von Barth}},
  \ and\ \bibinfo {author} {\bibfnamefont {R.}~\bibnamefont {van Leeuwen}},\
  }\href@noop {} {\bibfield  {journal} {\bibinfo  {journal} {Int. J. Mod. Phys.
  B}\ }\textbf {\bibinfo {volume} {13}},\ \bibinfo {pages} {535} (\bibinfo
  {year} {1999})}\BibitemShut {NoStop}%
\bibitem [{\citenamefont {Baym}(1962)}]{baym}%
  \BibitemOpen
  \bibfield  {author} {\bibinfo {author} {\bibfnamefont {G.}~\bibnamefont
  {Baym}},\ }\href {\doibase 10.1103/PhysRev.127.1391} {\bibfield  {journal}
  {\bibinfo  {journal} {Phys. Rev.}\ }\textbf {\bibinfo {volume} {127}},\
  \bibinfo {pages} {1391} (\bibinfo {year} {1962})}\BibitemShut {NoStop}%
\bibitem [{\citenamefont {Baym}\ and\ \citenamefont
  {Kadanoff}(1961)}]{baymkadanoff}%
  \BibitemOpen
  \bibfield  {author} {\bibinfo {author} {\bibfnamefont {G.}~\bibnamefont
  {Baym}}\ and\ \bibinfo {author} {\bibfnamefont {L.~P.}\ \bibnamefont
  {Kadanoff}},\ }\href {\doibase 10.1103/PhysRev.124.287} {\bibfield  {journal}
  {\bibinfo  {journal} {Phys. Rev.}\ }\textbf {\bibinfo {volume} {124}},\
  \bibinfo {pages} {287} (\bibinfo {year} {1961})}\BibitemShut {NoStop}%
\bibitem [{\citenamefont {Engel}(2003)}]{oep}%
  \BibitemOpen
  \bibfield  {author} {\bibinfo {author} {\bibfnamefont {E.}~\bibnamefont
  {Engel}},\ }in\ \href@noop {} {\emph {\bibinfo {booktitle} {A Primer in
  Density Functional Theory}}},\ \bibinfo {editor} {edited by\ \bibinfo
  {editor} {\bibfnamefont {C.~F.}\ \bibnamefont {et~al.}}}\ (\bibinfo
  {publisher} {Springer-Verlag Berlin Heidelberg},\ \bibinfo {year}
  {2003})\BibitemShut {NoStop}%
\bibitem [{\citenamefont {Del~Sole}\ \emph {et~al.}(1994)\citenamefont
  {Del~Sole}, \citenamefont {Reining},\ and\ \citenamefont
  {Godby}}]{PhysRevB.49.8024}%
  \BibitemOpen
  \bibfield  {author} {\bibinfo {author} {\bibfnamefont {R.}~\bibnamefont
  {Del~Sole}}, \bibinfo {author} {\bibfnamefont {L.}~\bibnamefont {Reining}}, \
  and\ \bibinfo {author} {\bibfnamefont {R.~W.}\ \bibnamefont {Godby}},\ }\href
  {\doibase 10.1103/PhysRevB.49.8024} {\bibfield  {journal} {\bibinfo
  {journal} {Phys. Rev. B}\ }\textbf {\bibinfo {volume} {49}},\ \bibinfo
  {pages} {8024} (\bibinfo {year} {1994})}\BibitemShut {NoStop}%
\bibitem [{\citenamefont {Bruneval}\ \emph {et~al.}(2005)\citenamefont
  {Bruneval}, \citenamefont {Sottile}, \citenamefont {Olevano}, \citenamefont
  {Del~Sole},\ and\ \citenamefont {Reining}}]{PhysRevLett.94.186402}%
  \BibitemOpen
  \bibfield  {author} {\bibinfo {author} {\bibfnamefont {F.}~\bibnamefont
  {Bruneval}}, \bibinfo {author} {\bibfnamefont {F.}~\bibnamefont {Sottile}},
  \bibinfo {author} {\bibfnamefont {V.}~\bibnamefont {Olevano}}, \bibinfo
  {author} {\bibfnamefont {R.}~\bibnamefont {Del~Sole}}, \ and\ \bibinfo
  {author} {\bibfnamefont {L.}~\bibnamefont {Reining}},\ }\href@noop {}
  {\bibfield  {journal} {\bibinfo  {journal} {Phys. Rev. Lett.}\ }\textbf
  {\bibinfo {volume} {94}},\ \bibinfo {pages} {186402} (\bibinfo {year}
  {2005})}\BibitemShut {NoStop}%
\bibitem [{\citenamefont {Romaniello}\ \emph {et~al.}(2009)\citenamefont
  {Romaniello}, \citenamefont {Guyot},\ and\ \citenamefont
  {Reining}}]{doi:10.1063/1.3249965}%
  \BibitemOpen
  \bibfield  {author} {\bibinfo {author} {\bibfnamefont {P.}~\bibnamefont
  {Romaniello}}, \bibinfo {author} {\bibfnamefont {S.}~\bibnamefont {Guyot}}, \
  and\ \bibinfo {author} {\bibfnamefont {L.}~\bibnamefont {Reining}},\ }\href
  {\doibase 10.1063/1.3249965} {\bibfield  {journal} {\bibinfo  {journal} {The
  Journal of Chemical Physics}\ }\textbf {\bibinfo {volume} {131}},\ \bibinfo
  {pages} {154111} (\bibinfo {year} {2009})}\BibitemShut {NoStop}%
\bibitem [{\citenamefont {Cao}\ \emph {et~al.}(2017)\citenamefont {Cao},
  \citenamefont {Yu}, \citenamefont {Lu},\ and\ \citenamefont
  {Wang}}]{PhysRevB.95.035139}%
  \BibitemOpen
  \bibfield  {author} {\bibinfo {author} {\bibfnamefont {H.}~\bibnamefont
  {Cao}}, \bibinfo {author} {\bibfnamefont {Z.}~\bibnamefont {Yu}}, \bibinfo
  {author} {\bibfnamefont {P.}~\bibnamefont {Lu}}, \ and\ \bibinfo {author}
  {\bibfnamefont {L.-W.}\ \bibnamefont {Wang}},\ }\href {\doibase
  10.1103/PhysRevB.95.035139} {\bibfield  {journal} {\bibinfo  {journal} {Phys.
  Rev. B}\ }\textbf {\bibinfo {volume} {95}},\ \bibinfo {pages} {035139}
  (\bibinfo {year} {2017})}\BibitemShut {NoStop}%
\bibitem [{\citenamefont {Toulouse}\ \emph {et~al.}(2009)\citenamefont
  {Toulouse}, \citenamefont {Gerber}, \citenamefont {Jansen}, \citenamefont
  {Savin},\ and\ \citenamefont {\'Angy\'an}}]{toulouse_prl}%
  \BibitemOpen
  \bibfield  {author} {\bibinfo {author} {\bibfnamefont {J.}~\bibnamefont
  {Toulouse}}, \bibinfo {author} {\bibfnamefont {I.~C.}\ \bibnamefont
  {Gerber}}, \bibinfo {author} {\bibfnamefont {G.}~\bibnamefont {Jansen}},
  \bibinfo {author} {\bibfnamefont {A.}~\bibnamefont {Savin}}, \ and\ \bibinfo
  {author} {\bibfnamefont {J.~G.}\ \bibnamefont {\'Angy\'an}},\ }\href
  {\doibase 10.1103/PhysRevLett.102.096404} {\bibfield  {journal} {\bibinfo
  {journal} {Phys. Rev. Lett.}\ }\textbf {\bibinfo {volume} {102}},\ \bibinfo
  {pages} {096404} (\bibinfo {year} {2009})}\BibitemShut {NoStop}%
\bibitem [{\citenamefont {Hellgren}\ and\ \citenamefont {von
  Barth}(2009)}]{hvb09}%
  \BibitemOpen
  \bibfield  {author} {\bibinfo {author} {\bibfnamefont {M.}~\bibnamefont
  {Hellgren}}\ and\ \bibinfo {author} {\bibfnamefont {U.}~\bibnamefont {von
  Barth}},\ }\href@noop {} {\bibfield  {journal} {\bibinfo  {journal} {J. Chem.
  Phys.}\ }\textbf {\bibinfo {volume} {131}},\ \bibinfo {pages} {044110}
  (\bibinfo {year} {2009})}\BibitemShut {NoStop}%
\bibitem [{\citenamefont {He\ss{}elmann}\ and\ \citenamefont
  {G\"orling}(2010)}]{hesselmann_random_2010}%
  \BibitemOpen
  \bibfield  {author} {\bibinfo {author} {\bibfnamefont {A.}~\bibnamefont
  {He\ss{}elmann}}\ and\ \bibinfo {author} {\bibfnamefont {A.}~\bibnamefont
  {G\"orling}},\ }\href {\doibase 10.1080/00268970903476662} {\bibfield
  {journal} {\bibinfo  {journal} {Molecular Physics}\ }\textbf {\bibinfo
  {volume} {108}},\ \bibinfo {pages} {359} (\bibinfo {year}
  {2010})}\BibitemShut {NoStop}%
\bibitem [{\citenamefont {Hellgren}\ and\ \citenamefont {von
  Barth}(2010)}]{hvb10}%
  \BibitemOpen
  \bibfield  {author} {\bibinfo {author} {\bibfnamefont {M.}~\bibnamefont
  {Hellgren}}\ and\ \bibinfo {author} {\bibfnamefont {U.}~\bibnamefont {von
  Barth}},\ }\href@noop {} {\bibfield  {journal} {\bibinfo  {journal} {J. Chem.
  Phys.}\ }\textbf {\bibinfo {volume} {132}},\ \bibinfo {pages} {044101}
  (\bibinfo {year} {2010})}\BibitemShut {NoStop}%
\bibitem [{\citenamefont {Hesselmann}\ and\ \citenamefont
  {G\"orling}(2011)}]{hg11}%
  \BibitemOpen
  \bibfield  {author} {\bibinfo {author} {\bibfnamefont {A.}~\bibnamefont
  {Hesselmann}}\ and\ \bibinfo {author} {\bibfnamefont {A.}~\bibnamefont
  {G\"orling}},\ }\href@noop {} {\bibfield  {journal} {\bibinfo  {journal}
  {Phys. Rev. Lett.}\ }\textbf {\bibinfo {volume} {106}},\ \bibinfo {pages}
  {093001} (\bibinfo {year} {2011})}\BibitemShut {NoStop}%
\bibitem [{\citenamefont {Bates}\ and\ \citenamefont {Furche}(2013)}]{RPAx-F}%
  \BibitemOpen
  \bibfield  {author} {\bibinfo {author} {\bibfnamefont {J.~E.}\ \bibnamefont
  {Bates}}\ and\ \bibinfo {author} {\bibfnamefont {F.}~\bibnamefont {Furche}},\
  }\href {\doibase http://dx.doi.org/10.1063/1.4827254} {\bibfield  {journal}
  {\bibinfo  {journal} {The Journal of Chemical Physics}\ }\textbf {\bibinfo
  {volume} {139}},\ \bibinfo {eid} {171103} (\bibinfo {year}
  {2013})}\BibitemShut {NoStop}%
\bibitem [{\citenamefont {Colonna}\ \emph {et~al.}(2014)\citenamefont
  {Colonna}, \citenamefont {Hellgren},\ and\ \citenamefont
  {de~Gironcoli}}]{PhysRevB.90.125150}%
  \BibitemOpen
  \bibfield  {author} {\bibinfo {author} {\bibfnamefont {N.}~\bibnamefont
  {Colonna}}, \bibinfo {author} {\bibfnamefont {M.}~\bibnamefont {Hellgren}}, \
  and\ \bibinfo {author} {\bibfnamefont {S.}~\bibnamefont {de~Gironcoli}},\
  }\href {\doibase 10.1103/PhysRevB.90.125150} {\bibfield  {journal} {\bibinfo
  {journal} {Phys. Rev. B}\ }\textbf {\bibinfo {volume} {90}},\ \bibinfo
  {pages} {125150} (\bibinfo {year} {2014})}\BibitemShut {NoStop}%
\bibitem [{\citenamefont {Dixit}\ \emph {et~al.}(2016)\citenamefont {Dixit},
  \citenamefont {{\'A}ngy{\'a}n},\ and\ \citenamefont
  {Rocca}}]{dixit_improving_2016}%
  \BibitemOpen
  \bibfield  {author} {\bibinfo {author} {\bibfnamefont {A.}~\bibnamefont
  {Dixit}}, \bibinfo {author} {\bibfnamefont {J.~G.}\ \bibnamefont
  {{\'A}ngy{\'a}n}}, \ and\ \bibinfo {author} {\bibfnamefont {D.}~\bibnamefont
  {Rocca}},\ }\href {\doibase 10.1063/1.4962352} {\bibfield  {journal}
  {\bibinfo  {journal} {The Journal of Chemical Physics}\ }\textbf {\bibinfo
  {volume} {145}},\ \bibinfo {pages} {104105} (\bibinfo {year}
  {2016})}\BibitemShut {NoStop}%
\bibitem [{\citenamefont {Mussard}\ \emph {et~al.}(2016)\citenamefont
  {Mussard}, \citenamefont {Rocca}, \citenamefont {Jansen},\ and\ \citenamefont
  {{\'A}ngy{\'a}n}}]{mussard_dielectric_2016}%
  \BibitemOpen
  \bibfield  {author} {\bibinfo {author} {\bibfnamefont {B.}~\bibnamefont
  {Mussard}}, \bibinfo {author} {\bibfnamefont {D.}~\bibnamefont {Rocca}},
  \bibinfo {author} {\bibfnamefont {G.}~\bibnamefont {Jansen}}, \ and\ \bibinfo
  {author} {\bibfnamefont {J.~G.}\ \bibnamefont {{\'A}ngy{\'a}n}},\ }\href
  {\doibase 10.1021/acs.jctc.5b01129} {\bibfield  {journal} {\bibinfo
  {journal} {J. Chem. Theory Comput.}\ }\textbf {\bibinfo {volume} {12}},\
  \bibinfo {pages} {2191} (\bibinfo {year} {2016})}\BibitemShut {NoStop}%
\bibitem [{\citenamefont {Dixit}\ \emph {et~al.}(2017)\citenamefont {Dixit},
  \citenamefont {Claudot}, \citenamefont {Lebègue},\ and\ \citenamefont
  {Rocca}}]{dixit17}%
  \BibitemOpen
  \bibfield  {author} {\bibinfo {author} {\bibfnamefont {A.}~\bibnamefont
  {Dixit}}, \bibinfo {author} {\bibfnamefont {J.}~\bibnamefont {Claudot}},
  \bibinfo {author} {\bibfnamefont {S.}~\bibnamefont {Lebegue}}, \ and\
  \bibinfo {author} {\bibfnamefont {D.}~\bibnamefont {Rocca}},\ }\href@noop {}
  {\bibfield  {journal} {\bibinfo  {journal} {Journal of Chemical Theory and
  Computation}\ }\textbf {\bibinfo {volume} {13}},\ \bibinfo {pages} {5432}
  (\bibinfo {year} {2017})}\BibitemShut {NoStop}%
\bibitem [{\citenamefont {{\v R}ez{\'a}{\v c}}\ and\ \citenamefont
  {Hobza}(2013)}]{rezac_describing_2013}%
  \BibitemOpen
  \bibfield  {author} {\bibinfo {author} {\bibfnamefont {J.}~\bibnamefont {{\v
  R}ez{\'a}{\v c}}}\ and\ \bibinfo {author} {\bibfnamefont {P.}~\bibnamefont
  {Hobza}},\ }\href {\doibase 10.1021/ct400057w} {\bibfield  {journal}
  {\bibinfo  {journal} {J. Chem. Theory Comput.}\ }\textbf {\bibinfo {volume}
  {9}},\ \bibinfo {pages} {2151} (\bibinfo {year} {2013})}\BibitemShut
  {NoStop}%
\bibitem [{\citenamefont {Fetter}\ and\ \citenamefont
  {Walecka}(2003)}]{fetter}%
  \BibitemOpen
  \bibfield  {author} {\bibinfo {author} {\bibfnamefont {A.~L.}\ \bibnamefont
  {Fetter}}\ and\ \bibinfo {author} {\bibfnamefont {J.~D.}\ \bibnamefont
  {Walecka}},\ }\href@noop {} {\emph {\bibinfo {title} {Quantum Theory of
  {Many-Particle} Systems}}}\ (\bibinfo  {publisher} {Dover Publications},\
  \bibinfo {year} {2003})\BibitemShut {NoStop}%
\bibitem [{\citenamefont {Langreth}\ and\ \citenamefont
  {Perdew}(1975)}]{langreth}%
  \BibitemOpen
  \bibfield  {author} {\bibinfo {author} {\bibfnamefont {D.}~\bibnamefont
  {Langreth}}\ and\ \bibinfo {author} {\bibfnamefont {J.}~\bibnamefont
  {Perdew}},\ }\href@noop {} {\bibfield  {journal} {\bibinfo  {journal} {Solid
  State Commun.}\ }\textbf {\bibinfo {volume} {17}},\ \bibinfo {pages} {1425}
  (\bibinfo {year} {1975})}\BibitemShut {NoStop}%
\bibitem [{\citenamefont {Caruso}\ \emph {et~al.}(2013)\citenamefont {Caruso},
  \citenamefont {Rohr}, \citenamefont {Hellgren}, \citenamefont {Ren},
  \citenamefont {Rinke}, \citenamefont {Rubio},\ and\ \citenamefont
  {Scheffler}}]{gwrpa1}%
  \BibitemOpen
  \bibfield  {author} {\bibinfo {author} {\bibfnamefont {F.}~\bibnamefont
  {Caruso}}, \bibinfo {author} {\bibfnamefont {D.~R.}\ \bibnamefont {Rohr}},
  \bibinfo {author} {\bibfnamefont {M.}~\bibnamefont {Hellgren}}, \bibinfo
  {author} {\bibfnamefont {X.}~\bibnamefont {Ren}}, \bibinfo {author}
  {\bibfnamefont {P.}~\bibnamefont {Rinke}}, \bibinfo {author} {\bibfnamefont
  {A.}~\bibnamefont {Rubio}}, \ and\ \bibinfo {author} {\bibfnamefont
  {M.}~\bibnamefont {Scheffler}},\ }\href {\doibase
  10.1103/PhysRevLett.110.146403} {\bibfield  {journal} {\bibinfo  {journal}
  {Physical Review Letters}\ }\textbf {\bibinfo {volume} {110}},\ \bibinfo
  {pages} {146403} (\bibinfo {year} {2013})}\BibitemShut {NoStop}%
\bibitem [{\citenamefont {Hellgren}\ \emph {et~al.}(2015)\citenamefont
  {Hellgren}, \citenamefont {Caruso}, \citenamefont {Rohr}, \citenamefont
  {Ren}, \citenamefont {Rubio}, \citenamefont {Scheffler},\ and\ \citenamefont
  {Rinke}}]{gwrpa2}%
  \BibitemOpen
  \bibfield  {author} {\bibinfo {author} {\bibfnamefont {M.}~\bibnamefont
  {Hellgren}}, \bibinfo {author} {\bibfnamefont {F.}~\bibnamefont {Caruso}},
  \bibinfo {author} {\bibfnamefont {D.~R.}\ \bibnamefont {Rohr}}, \bibinfo
  {author} {\bibfnamefont {X.}~\bibnamefont {Ren}}, \bibinfo {author}
  {\bibfnamefont {A.}~\bibnamefont {Rubio}}, \bibinfo {author} {\bibfnamefont
  {M.}~\bibnamefont {Scheffler}}, \ and\ \bibinfo {author} {\bibfnamefont
  {P.}~\bibnamefont {Rinke}},\ }\href {\doibase 10.1103/PhysRevB.91.165110}
  {\bibfield  {journal} {\bibinfo  {journal} {Physical Review B}\ }\textbf
  {\bibinfo {volume} {91}},\ \bibinfo {pages} {165110} (\bibinfo {year}
  {2015})}\BibitemShut {NoStop}%
\bibitem [{\citenamefont {Hellgren}\ and\ \citenamefont {von
  Barth}(2008)}]{hvb08}%
  \BibitemOpen
  \bibfield  {author} {\bibinfo {author} {\bibfnamefont {M.}~\bibnamefont
  {Hellgren}}\ and\ \bibinfo {author} {\bibfnamefont {U.}~\bibnamefont {von
  Barth}},\ }\href@noop {} {\bibfield  {journal} {\bibinfo  {journal} {Phys.
  Rev. B}\ }\textbf {\bibinfo {volume} {78}},\ \bibinfo {pages} {115107}
  (\bibinfo {year} {2008})}\BibitemShut {NoStop}%
\bibitem [{\citenamefont {Kim}\ and\ \citenamefont {G\"orling}(2002)}]{kg02}%
  \BibitemOpen
  \bibfield  {author} {\bibinfo {author} {\bibfnamefont {Y.-H.}\ \bibnamefont
  {Kim}}\ and\ \bibinfo {author} {\bibfnamefont {A.}~\bibnamefont
  {G\"orling}},\ }\href@noop {} {\bibfield  {journal} {\bibinfo  {journal}
  {Phys. Rev. B}\ }\textbf {\bibinfo {volume} {66}},\ \bibinfo {pages} {035144}
  (\bibinfo {year} {2002})}\BibitemShut {NoStop}%
\bibitem [{\citenamefont {Zhu}\ \emph {et~al.}(2010)\citenamefont {Zhu},
  \citenamefont {Toulouse}, \citenamefont {Savin},\ and\ \citenamefont
  {\'Angy\'an}}]{toulousejcp}%
  \BibitemOpen
  \bibfield  {author} {\bibinfo {author} {\bibfnamefont {W.}~\bibnamefont
  {Zhu}}, \bibinfo {author} {\bibfnamefont {J.}~\bibnamefont {Toulouse}},
  \bibinfo {author} {\bibfnamefont {A.}~\bibnamefont {Savin}}, \ and\ \bibinfo
  {author} {\bibfnamefont {J.~G.}\ \bibnamefont {\'Angy\'an}},\ }\href@noop {}
  {\bibfield  {journal} {\bibinfo  {journal} {The Journal of Chemical Physics}\
  }\textbf {\bibinfo {volume} {132}},\ \bibinfo {pages} {244108} (\bibinfo
  {year} {2010})}\BibitemShut {NoStop}%
\bibitem [{\citenamefont {Colonna}\ \emph {et~al.}(2016)\citenamefont
  {Colonna}, \citenamefont {Hellgren},\ and\ \citenamefont
  {de~Gironcoli}}]{PhysRevB.93.195108}%
  \BibitemOpen
  \bibfield  {author} {\bibinfo {author} {\bibfnamefont {N.}~\bibnamefont
  {Colonna}}, \bibinfo {author} {\bibfnamefont {M.}~\bibnamefont {Hellgren}}, \
  and\ \bibinfo {author} {\bibfnamefont {S.}~\bibnamefont {de~Gironcoli}},\
  }\href {\doibase 10.1103/PhysRevB.93.195108} {\bibfield  {journal} {\bibinfo
  {journal} {Phys. Rev. B}\ }\textbf {\bibinfo {volume} {93}},\ \bibinfo
  {pages} {195108} (\bibinfo {year} {2016})}\BibitemShut {NoStop}%
\bibitem [{\citenamefont {Giannozzi}\ \emph {et~al.}(2017)\citenamefont
  {Giannozzi}, \citenamefont {Andreussi}, \citenamefont {Brumme}, \citenamefont
  {Bunau}, \citenamefont {Nardelli}, \citenamefont {Calandra}, \citenamefont
  {Car}, \citenamefont {Cavazzoni}, \citenamefont {Ceresoli}, \citenamefont
  {Cococcioni}, \citenamefont {Colonna}, \citenamefont {Carnimeo},
  \citenamefont {Corso}, \citenamefont {de~Gironcoli}, \citenamefont {Delugas},
  \citenamefont {Jr}, \citenamefont {Ferretti}, \citenamefont {Floris},
  \citenamefont {Fratesi}, \citenamefont {Fugallo}, \citenamefont {Gebauer},
  \citenamefont {Gerstmann}, \citenamefont {Giustino}, \citenamefont {Gorni},
  \citenamefont {Jia}, \citenamefont {Kawamura}, \citenamefont {Ko},
  \citenamefont {Kokalj}, \citenamefont {c\'ukbenli}, \citenamefont {Lazzeri},
  \citenamefont {Marsili}, \citenamefont {Marzari}, \citenamefont {Mauri},
  \citenamefont {Nguyen}, \citenamefont {Nguyen}, \citenamefont {de-la Roza},
  \citenamefont {Paulatto}, \citenamefont {Ponce}, \citenamefont {Rocca},
  \citenamefont {Sabatini}, \citenamefont {Santra}, \citenamefont {Schlipf},
  \citenamefont {Seitsonen}, \citenamefont {Smogunov}, \citenamefont {Timrov},
  \citenamefont {Thonhauser}, \citenamefont {Umari}, \citenamefont {Vast},
  \citenamefont {Wu},\ and\ \citenamefont {Baroni}}]{giannozzi_advanced_2017}%
  \BibitemOpen
  \bibfield  {author} {\bibinfo {author} {\bibfnamefont {P.}~\bibnamefont
  {Giannozzi}}, \bibinfo {author} {\bibfnamefont {O.}~\bibnamefont
  {Andreussi}}, \bibinfo {author} {\bibfnamefont {T.}~\bibnamefont {Brumme}},
  \bibinfo {author} {\bibfnamefont {O.}~\bibnamefont {Bunau}}, \bibinfo
  {author} {\bibfnamefont {M.~B.}\ \bibnamefont {Nardelli}}, \bibinfo {author}
  {\bibfnamefont {M.}~\bibnamefont {Calandra}}, \bibinfo {author}
  {\bibfnamefont {R.}~\bibnamefont {Car}}, \bibinfo {author} {\bibfnamefont
  {C.}~\bibnamefont {Cavazzoni}}, \bibinfo {author} {\bibfnamefont
  {D.}~\bibnamefont {Ceresoli}}, \bibinfo {author} {\bibfnamefont
  {M.}~\bibnamefont {Cococcioni}}, \bibinfo {author} {\bibfnamefont
  {N.}~\bibnamefont {Colonna}}, \bibinfo {author} {\bibfnamefont
  {I.}~\bibnamefont {Carnimeo}}, \bibinfo {author} {\bibfnamefont {A.~D.}\
  \bibnamefont {Corso}}, \bibinfo {author} {\bibfnamefont {S.}~\bibnamefont
  {de~Gironcoli}}, \bibinfo {author} {\bibfnamefont {P.}~\bibnamefont
  {Delugas}}, \bibinfo {author} {\bibfnamefont {R.~A.~D.}\ \bibnamefont {Jr}},
  \bibinfo {author} {\bibfnamefont {A.}~\bibnamefont {Ferretti}}, \bibinfo
  {author} {\bibfnamefont {A.}~\bibnamefont {Floris}}, \bibinfo {author}
  {\bibfnamefont {G.}~\bibnamefont {Fratesi}}, \bibinfo {author} {\bibfnamefont
  {G.}~\bibnamefont {Fugallo}}, \bibinfo {author} {\bibfnamefont
  {R.}~\bibnamefont {Gebauer}}, \bibinfo {author} {\bibfnamefont
  {U.}~\bibnamefont {Gerstmann}}, \bibinfo {author} {\bibfnamefont
  {F.}~\bibnamefont {Giustino}}, \bibinfo {author} {\bibfnamefont
  {T.}~\bibnamefont {Gorni}}, \bibinfo {author} {\bibfnamefont
  {J.}~\bibnamefont {Jia}}, \bibinfo {author} {\bibfnamefont {M.}~\bibnamefont
  {Kawamura}}, \bibinfo {author} {\bibfnamefont {H.-Y.}\ \bibnamefont {Ko}},
  \bibinfo {author} {\bibfnamefont {A.}~\bibnamefont {Kokalj}}, \bibinfo
  {author} {\bibfnamefont {E.~E.~K.}\ \bibnamefont {c\'ukbenli}}, \bibinfo
  {author} {\bibfnamefont {M.}~\bibnamefont {Lazzeri}}, \bibinfo {author}
  {\bibfnamefont {M.}~\bibnamefont {Marsili}}, \bibinfo {author} {\bibfnamefont
  {N.}~\bibnamefont {Marzari}}, \bibinfo {author} {\bibfnamefont
  {F.}~\bibnamefont {Mauri}}, \bibinfo {author} {\bibfnamefont {N.~L.}\
  \bibnamefont {Nguyen}}, \bibinfo {author} {\bibfnamefont {H.-V.}\
  \bibnamefont {Nguyen}}, \bibinfo {author} {\bibfnamefont {A.~O.}\
  \bibnamefont {de-la Roza}}, \bibinfo {author} {\bibfnamefont
  {L.}~\bibnamefont {Paulatto}}, \bibinfo {author} {\bibfnamefont
  {S.}~\bibnamefont {Ponce}}, \bibinfo {author} {\bibfnamefont
  {D.}~\bibnamefont {Rocca}}, \bibinfo {author} {\bibfnamefont
  {R.}~\bibnamefont {Sabatini}}, \bibinfo {author} {\bibfnamefont
  {B.}~\bibnamefont {Santra}}, \bibinfo {author} {\bibfnamefont
  {M.}~\bibnamefont {Schlipf}}, \bibinfo {author} {\bibfnamefont {A.~P.}\
  \bibnamefont {Seitsonen}}, \bibinfo {author} {\bibfnamefont {A.}~\bibnamefont
  {Smogunov}}, \bibinfo {author} {\bibfnamefont {I.}~\bibnamefont {Timrov}},
  \bibinfo {author} {\bibfnamefont {T.}~\bibnamefont {Thonhauser}}, \bibinfo
  {author} {\bibfnamefont {P.}~\bibnamefont {Umari}}, \bibinfo {author}
  {\bibfnamefont {N.}~\bibnamefont {Vast}}, \bibinfo {author} {\bibfnamefont
  {X.}~\bibnamefont {Wu}}, \ and\ \bibinfo {author} {\bibfnamefont
  {S.}~\bibnamefont {Baroni}},\ }\href
  {http://stacks.iop.org/0953-8984/29/i=46/a=465901} {\bibfield  {journal}
  {\bibinfo  {journal} {Journal of Physics: Condensed Matter}\ }\textbf
  {\bibinfo {volume} {29}},\ \bibinfo {pages} {465901} (\bibinfo {year}
  {2017})}\BibitemShut {NoStop}%
\bibitem [{\citenamefont {Nguyen}\ and\ \citenamefont
  {de~Gironcoli}(2009)}]{nguyen_efficient_2009}%
  \BibitemOpen
  \bibfield  {author} {\bibinfo {author} {\bibfnamefont {H.-V.}\ \bibnamefont
  {Nguyen}}\ and\ \bibinfo {author} {\bibfnamefont {S.}~\bibnamefont
  {de~Gironcoli}},\ }\href {\doibase 10.1103/PhysRevB.79.205114} {\bibfield
  {journal} {\bibinfo  {journal} {Phys. Rev. B}\ }\textbf {\bibinfo {volume}
  {79}},\ \bibinfo {pages} {205114} (\bibinfo {year} {2009})}\BibitemShut
  {NoStop}%
\bibitem [{\citenamefont {Nguyen}\ \emph {et~al.}(2014)\citenamefont {Nguyen},
  \citenamefont {Colonna},\ and\ \citenamefont
  {de~Gironcoli}}]{nguyen_ab_2014}%
  \BibitemOpen
  \bibfield  {author} {\bibinfo {author} {\bibfnamefont {N.~L.}\ \bibnamefont
  {Nguyen}}, \bibinfo {author} {\bibfnamefont {N.}~\bibnamefont {Colonna}}, \
  and\ \bibinfo {author} {\bibfnamefont {S.}~\bibnamefont {de~Gironcoli}},\
  }\href {\doibase 10.1103/PhysRevB.90.045138} {\bibfield  {journal} {\bibinfo
  {journal} {Phys. Rev. B}\ }\textbf {\bibinfo {volume} {90}},\ \bibinfo
  {pages} {045138} (\bibinfo {year} {2014})}\BibitemShut {NoStop}%
\bibitem [{\citenamefont {Kolos}\ and\ \citenamefont
  {Wolniewics}(1965)}]{kolos}%
  \BibitemOpen
  \bibfield  {author} {\bibinfo {author} {\bibfnamefont {W.}~\bibnamefont
  {Kolos}}\ and\ \bibinfo {author} {\bibfnamefont {L.}~\bibnamefont
  {Wolniewics}},\ }\href@noop {} {\bibfield  {journal} {\bibinfo  {journal} {J.
  Chem. Phys.}\ }\textbf {\bibinfo {volume} {43}},\ \bibinfo {pages} {2429}
  (\bibinfo {year} {1965})}\BibitemShut {NoStop}%
\bibitem [{\citenamefont {Ren}\ \emph {et~al.}(2012)\citenamefont {Ren},
  \citenamefont {Rinke}, \citenamefont {Joas},\ and\ \citenamefont
  {Scheffler}}]{Ren2012}%
  \BibitemOpen
  \bibfield  {author} {\bibinfo {author} {\bibfnamefont {X.}~\bibnamefont
  {Ren}}, \bibinfo {author} {\bibfnamefont {P.}~\bibnamefont {Rinke}}, \bibinfo
  {author} {\bibfnamefont {C.}~\bibnamefont {Joas}}, \ and\ \bibinfo {author}
  {\bibfnamefont {M.}~\bibnamefont {Scheffler}},\ }\href {\doibase
  10.1007/s10853-012-6570-4} {\bibfield  {journal} {\bibinfo  {journal}
  {Journal of Materials Science}\ }\textbf {\bibinfo {volume} {47}},\ \bibinfo
  {pages} {7447} (\bibinfo {year} {2012})}\BibitemShut {NoStop}%
\bibitem [{\citenamefont {Ceperley}\ and\ \citenamefont
  {Alder}(1980)}]{ceperley_ground_1980}%
  \BibitemOpen
  \bibfield  {author} {\bibinfo {author} {\bibfnamefont {D.~M.}\ \bibnamefont
  {Ceperley}}\ and\ \bibinfo {author} {\bibfnamefont {B.~J.}\ \bibnamefont
  {Alder}},\ }\href {\doibase 10.1103/PhysRevLett.45.566} {\bibfield  {journal}
  {\bibinfo  {journal} {Phys. Rev. Lett.}\ }\textbf {\bibinfo {volume} {45}},\
  \bibinfo {pages} {566} (\bibinfo {year} {1980})}\BibitemShut {NoStop}%
\bibitem [{\citenamefont {Moroni}\ \emph {et~al.}(1995)\citenamefont {Moroni},
  \citenamefont {Ceperley},\ and\ \citenamefont {Senatore}}]{egqmc}%
  \BibitemOpen
  \bibfield  {author} {\bibinfo {author} {\bibfnamefont {S.}~\bibnamefont
  {Moroni}}, \bibinfo {author} {\bibfnamefont {D.~M.}\ \bibnamefont
  {Ceperley}}, \ and\ \bibinfo {author} {\bibfnamefont {G.}~\bibnamefont
  {Senatore}},\ }\href@noop {} {\bibfield  {journal} {\bibinfo  {journal}
  {Phys. Rev. Lett.}\ }\textbf {\bibinfo {volume} {75}},\ \bibinfo {pages}
  {689} (\bibinfo {year} {1995})}\BibitemShut {NoStop}%
\bibitem [{\citenamefont {Maggio}\ and\ \citenamefont
  {Kresse}(2016)}]{maggio_correlation_2016}%
  \BibitemOpen
  \bibfield  {author} {\bibinfo {author} {\bibfnamefont {E.}~\bibnamefont
  {Maggio}}\ and\ \bibinfo {author} {\bibfnamefont {G.}~\bibnamefont
  {Kresse}},\ }\href {\doibase 10.1103/PhysRevB.93.235113} {\bibfield
  {journal} {\bibinfo  {journal} {Phys. Rev. B}\ }\textbf {\bibinfo {volume}
  {93}},\ \bibinfo {pages} {235113} (\bibinfo {year} {2016})}\BibitemShut
  {NoStop}%
\bibitem [{\citenamefont {Toulouse}\ \emph {et~al.}(2011)\citenamefont
  {Toulouse}, \citenamefont {Zhu}, \citenamefont {Savin}, \citenamefont
  {Jansen},\ and\ \citenamefont {\'Angy\'an}}]{toulousejcp1}%
  \BibitemOpen
  \bibfield  {author} {\bibinfo {author} {\bibfnamefont {J.}~\bibnamefont
  {Toulouse}}, \bibinfo {author} {\bibfnamefont {W.}~\bibnamefont {Zhu}},
  \bibinfo {author} {\bibfnamefont {A.}~\bibnamefont {Savin}}, \bibinfo
  {author} {\bibfnamefont {G.}~\bibnamefont {Jansen}}, \ and\ \bibinfo {author}
  {\bibfnamefont {J.~G.}\ \bibnamefont {\'Angy\'an}},\ }\href@noop {}
  {\bibfield  {journal} {\bibinfo  {journal} {The Journal of Chemical Physics}\
  }\textbf {\bibinfo {volume} {135}},\ \bibinfo {pages} {084119} (\bibinfo
  {year} {2011})}\BibitemShut {NoStop}%
\bibitem [{\citenamefont {Erhard}\ \emph {et~al.}(2016)\citenamefont {Erhard},
  \citenamefont {Bleiziffer},\ and\ \citenamefont {G\"orling}}]{prlerhard}%
  \BibitemOpen
  \bibfield  {author} {\bibinfo {author} {\bibfnamefont {J.}~\bibnamefont
  {Erhard}}, \bibinfo {author} {\bibfnamefont {P.}~\bibnamefont {Bleiziffer}},
  \ and\ \bibinfo {author} {\bibfnamefont {A.}~\bibnamefont {G\"orling}},\
  }\href {\doibase 10.1103/PhysRevLett.117.143002} {\bibfield  {journal}
  {\bibinfo  {journal} {Phys. Rev. Lett.}\ }\textbf {\bibinfo {volume} {117}},\
  \bibinfo {pages} {143002} (\bibinfo {year} {2016})}\BibitemShut {NoStop}%
\bibitem [{\citenamefont {Giannozzi}\ \emph {et~al.}(2009)\citenamefont
  {Giannozzi}, \citenamefont {Baroni}, \citenamefont {Bonini}, \citenamefont
  {Calandra}, \citenamefont {Car}, \citenamefont {Cavazzoni}, \citenamefont
  {Ceresoli}, \citenamefont {Chiarotti}, \citenamefont {Cococcioni},
  \citenamefont {Dabo}, \citenamefont {Corso}, \citenamefont {Gironcoli},
  \citenamefont {Fabris}, \citenamefont {Fratesi}, \citenamefont {Gebauer},
  \citenamefont {Gerstmann}, \citenamefont {Gougoussis}, \citenamefont
  {Kokalj}, \citenamefont {Lazzeri}, \citenamefont {Martin-Samos},
  \citenamefont {Marzari}, \citenamefont {Mauri}, \citenamefont {Mazzarello},
  \citenamefont {Paolini}, \citenamefont {Pasquarello}, \citenamefont
  {Paulatto}, \citenamefont {Sbraccia}, \citenamefont {Scandolo}, \citenamefont
  {Sclauzero}, \citenamefont {Seitsonen}, \citenamefont {Smogunov},
  \citenamefont {Umari},\ and\ \citenamefont
  {Wentzcovitch}}]{giannozzi_quantum_2009}%
  \BibitemOpen
  \bibfield  {author} {\bibinfo {author} {\bibfnamefont {P.}~\bibnamefont
  {Giannozzi}}, \bibinfo {author} {\bibfnamefont {S.}~\bibnamefont {Baroni}},
  \bibinfo {author} {\bibfnamefont {N.}~\bibnamefont {Bonini}}, \bibinfo
  {author} {\bibfnamefont {M.}~\bibnamefont {Calandra}}, \bibinfo {author}
  {\bibfnamefont {R.}~\bibnamefont {Car}}, \bibinfo {author} {\bibfnamefont
  {C.}~\bibnamefont {Cavazzoni}}, \bibinfo {author} {\bibfnamefont
  {D.}~\bibnamefont {Ceresoli}}, \bibinfo {author} {\bibfnamefont {G.~L.}\
  \bibnamefont {Chiarotti}}, \bibinfo {author} {\bibfnamefont {M.}~\bibnamefont
  {Cococcioni}}, \bibinfo {author} {\bibfnamefont {I.}~\bibnamefont {Dabo}},
  \bibinfo {author} {\bibfnamefont {A.~D.}\ \bibnamefont {Corso}}, \bibinfo
  {author} {\bibfnamefont {S.~d.}\ \bibnamefont {Gironcoli}}, \bibinfo {author}
  {\bibfnamefont {S.}~\bibnamefont {Fabris}}, \bibinfo {author} {\bibfnamefont
  {G.}~\bibnamefont {Fratesi}}, \bibinfo {author} {\bibfnamefont
  {R.}~\bibnamefont {Gebauer}}, \bibinfo {author} {\bibfnamefont
  {U.}~\bibnamefont {Gerstmann}}, \bibinfo {author} {\bibfnamefont
  {C.}~\bibnamefont {Gougoussis}}, \bibinfo {author} {\bibfnamefont
  {A.}~\bibnamefont {Kokalj}}, \bibinfo {author} {\bibfnamefont
  {M.}~\bibnamefont {Lazzeri}}, \bibinfo {author} {\bibfnamefont
  {L.}~\bibnamefont {Martin-Samos}}, \bibinfo {author} {\bibfnamefont
  {N.}~\bibnamefont {Marzari}}, \bibinfo {author} {\bibfnamefont
  {F.}~\bibnamefont {Mauri}}, \bibinfo {author} {\bibfnamefont
  {R.}~\bibnamefont {Mazzarello}}, \bibinfo {author} {\bibfnamefont
  {S.}~\bibnamefont {Paolini}}, \bibinfo {author} {\bibfnamefont
  {A.}~\bibnamefont {Pasquarello}}, \bibinfo {author} {\bibfnamefont
  {L.}~\bibnamefont {Paulatto}}, \bibinfo {author} {\bibfnamefont
  {C.}~\bibnamefont {Sbraccia}}, \bibinfo {author} {\bibfnamefont
  {S.}~\bibnamefont {Scandolo}}, \bibinfo {author} {\bibfnamefont
  {G.}~\bibnamefont {Sclauzero}}, \bibinfo {author} {\bibfnamefont {A.~P.}\
  \bibnamefont {Seitsonen}}, \bibinfo {author} {\bibfnamefont {A.}~\bibnamefont
  {Smogunov}}, \bibinfo {author} {\bibfnamefont {P.}~\bibnamefont {Umari}}, \
  and\ \bibinfo {author} {\bibfnamefont {R.~M.}\ \bibnamefont {Wentzcovitch}},\
  }\href {\doibase 10.1088/0953-8984/21/39/395502} {\bibfield  {journal}
  {\bibinfo  {journal} {J. Phys.: Condens. Matter}\ }\textbf {\bibinfo {volume}
  {21}},\ \bibinfo {pages} {395502} (\bibinfo {year} {2009})}\BibitemShut
  {NoStop}%
\bibitem [{\citenamefont {Perdew}\ \emph {et~al.}(1996)\citenamefont {Perdew},
  \citenamefont {Burke},\ and\ \citenamefont {Ernzerhof}}]{pbe}%
  \BibitemOpen
  \bibfield  {author} {\bibinfo {author} {\bibfnamefont {J.~P.}\ \bibnamefont
  {Perdew}}, \bibinfo {author} {\bibfnamefont {K.}~\bibnamefont {Burke}}, \
  and\ \bibinfo {author} {\bibfnamefont {M.}~\bibnamefont {Ernzerhof}},\ }\href
  {\doibase 10.1103/PhysRevLett.77.3865} {\bibfield  {journal} {\bibinfo
  {journal} {Phys. Rev. Lett.}\ }\textbf {\bibinfo {volume} {77}},\ \bibinfo
  {pages} {3865} (\bibinfo {year} {1996})}\BibitemShut {NoStop}%
\bibitem [{\citenamefont {Hamann}(2013)}]{hamann_optimized_2013}%
  \BibitemOpen
  \bibfield  {author} {\bibinfo {author} {\bibfnamefont {D.~R.}\ \bibnamefont
  {Hamann}},\ }\href {\doibase 10.1103/PhysRevB.88.085117} {\bibfield
  {journal} {\bibinfo  {journal} {Phys. Rev. B}\ }\textbf {\bibinfo {volume}
  {88}},\ \bibinfo {pages} {085117} (\bibinfo {year} {2013})}\BibitemShut
  {NoStop}%
\bibitem [{\citenamefont {Schlipf}\ and\ \citenamefont
  {Gygi}(2015)}]{schlipf_optimization_2015}%
  \BibitemOpen
  \bibfield  {author} {\bibinfo {author} {\bibfnamefont {M.}~\bibnamefont
  {Schlipf}}\ and\ \bibinfo {author} {\bibfnamefont {F.}~\bibnamefont {Gygi}},\
  }\href {\doibase 10.1016/j.cpc.2015.05.011} {\bibfield  {journal} {\bibinfo
  {journal} {Computer Physics Communications}\ }\textbf {\bibinfo {volume}
  {196}},\ \bibinfo {pages} {36} (\bibinfo {year} {2015})}\BibitemShut
  {NoStop}%
\bibitem [{ONC(2017)}]{ONCV_website}%
  \BibitemOpen
  \href {http://www.quantum-simulation.org/potentials/sg15_oncv/} {\enquote
  {\bibinfo {title} {{SG15} {ONCV} {Potentials}},}\ } (\bibinfo {year}
  {2017})\BibitemShut {NoStop}%
\bibitem [{\citenamefont {Gygi}\ and\ \citenamefont
  {Baldereschi}(1986)}]{gygi_self-consistent_1986}%
  \BibitemOpen
  \bibfield  {author} {\bibinfo {author} {\bibfnamefont {F.}~\bibnamefont
  {Gygi}}\ and\ \bibinfo {author} {\bibfnamefont {A.}~\bibnamefont
  {Baldereschi}},\ }\href {\doibase 10.1103/PhysRevB.34.4405} {\bibfield
  {journal} {\bibinfo  {journal} {Phys. Rev. B}\ }\textbf {\bibinfo {volume}
  {34}},\ \bibinfo {pages} {4405} (\bibinfo {year} {1986})}\BibitemShut
  {NoStop}%
\bibitem [{\citenamefont {Spencer}\ and\ \citenamefont
  {Alavi}(2008)}]{spencer_efficient_2008}%
  \BibitemOpen
  \bibfield  {author} {\bibinfo {author} {\bibfnamefont {J.}~\bibnamefont
  {Spencer}}\ and\ \bibinfo {author} {\bibfnamefont {A.}~\bibnamefont
  {Alavi}},\ }\href {\doibase 10.1103/PhysRevB.77.193110} {\bibfield  {journal}
  {\bibinfo  {journal} {Phys. Rev. B}\ }\textbf {\bibinfo {volume} {77}},\
  \bibinfo {pages} {193110} (\bibinfo {year} {2008})}\BibitemShut {NoStop}%
\bibitem [{\citenamefont {Marsili}\ and\ \citenamefont
  {Umari}(2013)}]{marsili_method_2013}%
  \BibitemOpen
  \bibfield  {author} {\bibinfo {author} {\bibfnamefont {M.}~\bibnamefont
  {Marsili}}\ and\ \bibinfo {author} {\bibfnamefont {P.}~\bibnamefont
  {Umari}},\ }\href {\doibase 10.1103/PhysRevB.87.205110} {\bibfield  {journal}
  {\bibinfo  {journal} {Phys. Rev. B}\ }\textbf {\bibinfo {volume} {87}},\
  \bibinfo {pages} {205110} (\bibinfo {year} {2013})}\BibitemShut {NoStop}%
\bibitem [{\citenamefont {Rezac}\ and\ \citenamefont {Hobza}(2013)}]{refA24}%
  \BibitemOpen
  \bibfield  {author} {\bibinfo {author} {\bibfnamefont {J.}~\bibnamefont
  {Rezac}}\ and\ \bibinfo {author} {\bibfnamefont {P.}~\bibnamefont {Hobza}},\
  }\href@noop {} {\bibfield  {journal} {\bibinfo  {journal} {J. Chem. Theo.
  Comp.}\ }\textbf {\bibinfo {volume} {9}},\ \bibinfo {pages} {2151} (\bibinfo
  {year} {2013})}\BibitemShut {NoStop}%
\bibitem [{\citenamefont {Baroni}\ \emph {et~al.}(2001)\citenamefont {Baroni},
  \citenamefont {de~Gironcoli}, \citenamefont {Dal~Corso},\ and\ \citenamefont
  {Giannozzi}}]{baroni_phonons_2001}%
  \BibitemOpen
  \bibfield  {author} {\bibinfo {author} {\bibfnamefont {S.}~\bibnamefont
  {Baroni}}, \bibinfo {author} {\bibfnamefont {S.}~\bibnamefont
  {de~Gironcoli}}, \bibinfo {author} {\bibfnamefont {A.}~\bibnamefont
  {Dal~Corso}}, \ and\ \bibinfo {author} {\bibfnamefont {P.}~\bibnamefont
  {Giannozzi}},\ }\href {\doibase 10.1103/RevModPhys.73.515} {\bibfield
  {journal} {\bibinfo  {journal} {Rev. Mod. Phys.}\ }\textbf {\bibinfo {volume}
  {73}},\ \bibinfo {pages} {515} (\bibinfo {year} {2001})}\BibitemShut
  {NoStop}%
\bibitem [{Note1()}]{Note1}%
  \BibitemOpen
  \bibinfo {note} {This can be done since that the density\IeC {\textendash
  }density response functions are analytic in the upper half-plane and vanish
  at infinity faster than $\omega ^{-1}$ (see for instance Ref.~\protect
  \rev@citealpnum {giuliani_quantum_2005}).}\BibitemShut {Stop}%
\bibitem [{\citenamefont {Baldereschi}\ and\ \citenamefont
  {Tosatti}(1979)}]{baldereschi_dielectric_1979}%
  \BibitemOpen
  \bibfield  {author} {\bibinfo {author} {\bibfnamefont {A.}~\bibnamefont
  {Baldereschi}}\ and\ \bibinfo {author} {\bibfnamefont {E.}~\bibnamefont
  {Tosatti}},\ }\href {\doibase 10.1016/0038-1098(79)91022-6} {\bibfield
  {journal} {\bibinfo  {journal} {Solid State Communications}\ }\textbf
  {\bibinfo {volume} {29}},\ \bibinfo {pages} {131} (\bibinfo {year}
  {1979})}\BibitemShut {NoStop}%
\bibitem [{\citenamefont {Wilson}\ \emph {et~al.}(2008)\citenamefont {Wilson},
  \citenamefont {Gygi},\ and\ \citenamefont {Galli}}]{wilson_efficient_2008}%
  \BibitemOpen
  \bibfield  {author} {\bibinfo {author} {\bibfnamefont {H.~F.}\ \bibnamefont
  {Wilson}}, \bibinfo {author} {\bibfnamefont {F.}~\bibnamefont {Gygi}}, \ and\
  \bibinfo {author} {\bibfnamefont {G.}~\bibnamefont {Galli}},\ }\href
  {\doibase 10.1103/PhysRevB.78.113303} {\bibfield  {journal} {\bibinfo
  {journal} {Phys. Rev. B}\ }\textbf {\bibinfo {volume} {78}},\ \bibinfo
  {pages} {113303} (\bibinfo {year} {2008})}\BibitemShut {NoStop}%
\bibitem [{\citenamefont {Nguyen}(2008)}]{viet_phd_thesis}%
  \BibitemOpen
  \bibfield  {author} {\bibinfo {author} {\bibfnamefont {H.-V.}\ \bibnamefont
  {Nguyen}},\ }\emph {\bibinfo {title} {Efficient calculation of RPA
  correlation energies in the Adiabatic-Connection Fluctuation-Dissipation
  theory}},\ \href {http://preprints.sissa.it/xmlui/handle/1963/5260} {Ph.D.
  thesis},\ \bibinfo  {school} {Internatinal School for Advanced Studies
  (SISSA)} (\bibinfo {year} {2008})\BibitemShut {NoStop}%
\bibitem [{\citenamefont
  {G{\"o}rling}(1998{\natexlab{a}})}]{gorling_exact_1998_pra}%
  \BibitemOpen
  \bibfield  {author} {\bibinfo {author} {\bibfnamefont {A.}~\bibnamefont
  {G{\"o}rling}},\ }\href {\doibase 10.1103/PhysRevA.57.3433} {\bibfield
  {journal} {\bibinfo  {journal} {Phys. Rev. A}\ }\textbf {\bibinfo {volume}
  {57}},\ \bibinfo {pages} {3433} (\bibinfo {year}
  {1998}{\natexlab{a}})}\BibitemShut {NoStop}%
\bibitem [{\citenamefont
  {G{\"o}rling}(1998{\natexlab{b}})}]{gorling_exact_1998}%
  \BibitemOpen
  \bibfield  {author} {\bibinfo {author} {\bibfnamefont {A.}~\bibnamefont
  {G{\"o}rling}},\ }\href {\doibase
  10.1002/(SICI)1097-461X(1998)69:3<265::AID-QUA6>3.0.CO;2-T} {\bibfield
  {journal} {\bibinfo  {journal} {Int. J. Quantum Chem.}\ }\textbf {\bibinfo
  {volume} {69}},\ \bibinfo {pages} {265} (\bibinfo {year}
  {1998}{\natexlab{b}})}\BibitemShut {NoStop}%
\bibitem [{\citenamefont {Colonna}(2014)}]{NsC_phd_thesis}%
  \BibitemOpen
  \bibfield  {author} {\bibinfo {author} {\bibfnamefont {N.}~\bibnamefont
  {Colonna}},\ }\emph {\bibinfo {title} {Exchange and Correlation Energy in the
  Adiabatic Connection Fluctuation-Dissipation theory beyond RPA}},\ \href
  {http://preprints.sissa.it/xmlui/handle/1963/7468} {Ph.D. thesis},\ \bibinfo
  {school} {Internatinal School for Advanced Studies (SISSA)} (\bibinfo {year}
  {2014})\BibitemShut {NoStop}%
\bibitem [{\citenamefont {Giuliani}\ and\ \citenamefont
  {Vignale}(2005)}]{giuliani_quantum_2005}%
  \BibitemOpen
  \bibfield  {author} {\bibinfo {author} {\bibfnamefont {G.}~\bibnamefont
  {Giuliani}}\ and\ \bibinfo {author} {\bibfnamefont {G.}~\bibnamefont
  {Vignale}},\ }\href@noop {} {\emph {\bibinfo {title} {Quantum Theory of
  Electron Liquid}}}\ (\bibinfo  {publisher} {Cambridge University Press},\
  \bibinfo {year} {2005})\BibitemShut {NoStop}%
\end{thebibliography}
\end{document}